\documentstyle[12pt]{article}
\input epsf
\parskip=\medskipamount
\def\al{\alpha}
\def\be{\beta}
\def\ga{\gamma}

\def\te{\theta}   
   
\def\p{\phi}     
   
\def\IC{\relax{\rm l\kern-.50 em C}}
\def\IE{\relax{\rm l\kern-.12 em E}}
\def\IK{\relax{\rm l\kern-.18 em K}}
\def\IL{\relax{\rm I\kern-.18 em L}}
\def\IN{\relax{\rm I\kern-.18 em N}}
\def\IR{\relax{\rm I\kern-.18 em R}}

\font\tenfrak=eufm10  \font\sevenfrak=eufm7  \font\fivefrak=eufm5
\newfam\frakfam
\textfont\frakfam=\tenfrak
\scriptfont\frakfam=\sevenfrak
\scriptscriptfont\frakfam=\fivefrak

\def\tg{\mathop{\rm tg}\nolimits}

\def\wt{\widetilde}
\def\frac#1#2{{#1\over #2}}
\def\fracpd#1#2{\frac{\partial #1}{\partial #2}}
\def\Cos{\mathop{\rm C}\nolimits}    
\def\Sin{\mathop{\rm S}\nolimits}    
\def\Tan{\mathop{\rm T}\nolimits}    
\def\k{\kappa}                       

\def\f{\varphi}
\def\e{e}              

\begin{document}

\title{ Central potentials on spaces of constant curvature:
The Kepler problem on the two-dimensional sphere $S^2$ and
the hyperbolic plane $H^2$ }
\author{Jos\'e F. Cari\~nena$\dagger\,^{a)}$,
         Manuel F. Ra\~nada$\dagger\,^{b)}$  and
         Mariano Santander$\ddagger\,^{c)}$ \\[4pt]
$\dagger${\enskip}
   {\sl Departamento de F\'{\i}sica Te\'orica, Facultad de Ciencias}\\
   {\sl Universidad de Zaragoza, 50009 Zaragoza, Spain}  \\[2pt]
$\ddagger${\enskip}
   {\sl Departamento de F\'{\i}sica Te\'orica, Facultad de Ciencias}\\
   {\sl Universidad de Valladolid,  47011 Valladolid, Spain} }
\date{}
\maketitle

\begin{abstract}
The Kepler problem is a dynamical system that is well defined
not only on the Euclidean plane but also on the sphere and on
the hyperbolic plane.
First, the theory of central potentials on spaces of constant curvature
is studied.
All the mathematical expressions are presented using the curvature
$\k$ as a parameter, in such a way that they reduce to the appropriate
property for the system on the sphere $S^2$, or on the hyperbolic plane $H^2$,
when particularized for $\k>0$, or $\k<0$, respectively; in addition,
the Euclidean case arises as the particular case $\k=0$.
In the second part we study the main properties of the Kepler
problem on spaces with curvature, we solve the equations and we obtain
the explicit expressions of the orbits by using two different methods:
first by direct integration and second by
obtaining the $\k$-dependent version of the Binet's equation.
The final part of the article, that has a more geometric character,
is devoted to the study of the theory of conics on spaces of constant
curvature.
\end{abstract}

\begin{quote}
Kepler problem.  Integrability and Superintegrability.
Conics on the sphere. Conics on the hyperbolic plane.

{\sl Running title:}{\enskip}
The Kepler problem on spaces with constant curvature.


\end{quote}
{\vfill}

\footnoterule
{\noindent\small
$^{a)}${\sl E-mail address:} {jfc@unizar.es} \\
$^{b)}${\sl E-mail address:} {mfran@unizar.es} \\
$^{c)}${\sl E-mail address:} {santander@fta.uva.es} }
\newpage

\section{Introduction}

   The hydrogen atom in a spherical geometry $S^3$ was firstly studied
by Schr\"odinger \cite{Sch40} in 1940 and analyzed by Infeld \cite{In41} and
Stevenson \cite{St41} at the next year;
some five years later Infeld and Schild \cite{IS45} obtained the spectrum
of this system in an ``open universe of constant negative curvature"
(hyperbolic space $H^3$).
Later on, Higgs \cite{Hi79} and Leemon \cite{Le79} analyzed the characteristics
of the two fundamental central potentials, Kepler problem and 
harmonic oscillator,
on the $N$-dimensional sphere
and Kalnins {\sl et al}, in a study of the dynamical $O(4)$ symmetry of
the hydrogen atom \cite{KaMiW76}, considered the separation of variables of
the Schr\"odinger equation with a Coulomb potentialn on the sphere $S^3$
using a classifications of coordinate systems separating the Laplace equation
in the Riemannian  spaces of constant curvature previously obtained 
by Olevskii \cite{Olev50}.
Since then, a certain number of authors have studied this question from
both the classical (Kepler problem) and the quantum (hydrogen atom)
points of view \cite{BaW85}--\cite{RaS03}.
The Kepler problem and the harmonic oscillator are the two most important
superseparable systems and, because of this, their spherical and hyperbolic
versions have been occasionally obtained, in some of the quoted references,
as particular cases in the study of superintegrable systems on non-Euclidean
configuration spaces.

    The Euclidean superintegrable systems in the plane were studied by Fris
{\sl et al} \cite{FrMS65} and in the three-dimensional space by
Evans \cite{Ev90}; later on different authors have considered this
question \cite{GrPS95a}--\cite{GrW02} from different points of
view. In Ref.\ \cite{RaS99} we studied the existence of the
spherical and hyperbolic versions of these superintegrable systems
taking the space curvature as a parameter; next the properties of
the isotropic and nonisotropic harmonic oscillator and of those
superintegrable systems related with the harmonic oscillator were
analyzed \cite{RaS02a,RaS02b,RaSS02,RaS03}. Now, our objective is
the study of the Kepler problem on the sphere $S^2$ and the
hyperbolic plane $H^2$, both coming naturally from reduction from
the `true' Kepler problem in a three-dimensional curved space
$S^3$ or $H^3$.

   In differential geometric terms, the Euclidean plane $\IE^2$ is in fact
a very particular `limiting  case' of the constant curvature spaces;
accordingly, in dynamical terms, certain classical and well known potentials
(Kepler problem, harmonic oscillator, and systems related with them)
can also be considered as very particular `limiting  cases' of more general
`curved' systems.
If the curvature $\k$ is introduced as a parameter then the question
is the analysis of certain potentials in the space with constant curvature
$\k$ (we may term these as $\k$-dependent potentials) with appropriate
flat limit.
As many different $\k$-dependent functions may have the same limit when
$\k\to 0$ we must require that certain fundamental properties of the
Euclidean system continue to hold for the `curved' system.
By fundamental properties we mean those related with separability
(superseparability) and integrability (superintegrability).
The important point is that the spherical and hyperbolic Kepler
potential, to be studied in this article, can be characterized as a
very specific $\k$-deformation of the well known Euclidean system,
and conversely, the Euclidean Kepler potential arises as the particular
$\k=0$ case of the $\k$-dependent `curved' system.

   The main purpose of this article is to study and solve the Kepler problem
in the space of constant curvature $\k$. As we will see, one of the advantages
of this $\k$-dependent formalism is that the study runs parallel to that
of the Euclidean case. So the structure of this article is very similar
to that of the corresponding chapter (or sections) devoted to central
potentials and Kepler problem in some classical books of theoretical
mechanics \cite{Wh65,Go80}.
In fact this is the idea of using the deformation as an approach:
the curvature $\k$ will modify many things but will preserve the
fundamental structure; this is so because the deep properties ensuring
the possibility of exact solution for the Euclidean Kepler problem are
preserved when a constant curvature is ascribed to the space.
Of course, there are significant differences; one of them will be the
geometric interpretation of the solutions that, although rather simple
in the Euclidean plane, will pose some interesting geometric questions
in spaces of constant curvature.

  In more detail, the plan of this article is as follows:
In Sec.\ II we analyze the free geodesic motion on the spaces
$S^2$, $\IE^2$, and $H^2$, as well as the existence of
$\k$-dependent Noether symmetries.
In Sec.\ III we study some general questions concerning the
central potentials on spaces of constant curvature.
Sec.\ IV, that can be considered as the central part of the
article, is devoted to the study the $\k$-dependent Kepler problem
on the spaces $(S^2,\IE^2,H^2)$.
We have divided this section in three subsections:
in the first part we consider the equivalent one--dimensional problem
and we draw a rough classification of the orbits without requiring
explicit solutions;
in the second part we solve explicitly the problem and we obtain the
expressions of the orbits; we arrive to the results by using two different
methods: firstly by direct integration and secondly by
obtaining the $\k$-dependent version of the Binet's equation.
The third subsection is devoted to the analysis and classification
of the $\k$-dependent orbits; first on the sphere $S^2$ ($\k>0$),
and then in the hyperbolic plane $H^2$ ($\k<0$).
This third subsection poses the difficult problem of interpreting
the $\k$-dependent equations as conics in curved spaces; of course
all the obtained results reduce to well known Euclidean trajectories
when we consider the particular value $\k=0$.
Sec.\ V, that has a geometric character, is devoted to the study of
(non-Euclidean) conics on the sphere $S^2$ and on the hyperbolic
plane $H^2$, which leads to the identification of the orbits of the
Kepler problem with conics for any value of the curvature.
Finally, in Sec. VI we discuss the results and make some final comments.

\section{Geodesic motion, Noether symmetries and Constants of
Motion on $(S^2,\IE^2,H^2)$}

   In the following we will make use of the same notation and techniques
introduced for the oscillator in the above mentioned previous articles.
That is, from now on all the mathematical expressions will depend of
the curvature $\k$ as a parameter, in such a way that when assuming $\k>0$,
$\k=0$, or $\k<0$, we will obtain the corresponding property
particularized on the sphere, on the Euclidean plane, or on the
hyperbolic plane.
In order to present these expressions in a form which holds
simultaneously for any value of $\k$, we will use the following
``tagged" trigonometric functions
\begin{equation}
   \Cos_{\k}(x) = \cases{
   \cos{\sqrt{\k}\,x}       &if $\k>0$, \cr
   {\quad}  1               &if $\k=0$, \cr
   \cosh\!{\sqrt{-\k}\,x}   &if $\k<0$, \cr}{\qquad}
   \Sin_{\k}(x) = \cases{
   \frac{1}{\sqrt{\k}} \sin{\sqrt{\k}\,x}     &if $\k>0$, \cr
   {\quad}   x                                &if $\k=0$, \cr
   \frac{1}{\sqrt{-\k}}\sinh\!{\sqrt{-\k}\,x} &if $\k<0$, \label{SkCk}\cr}
\end{equation}
and the $\k$-dependent tangent function $\Tan_\k(x)$ defined in the 
natural way,
$\Tan_\k(x) = \Sin_{\k}(x)/\Cos_{\k}(x)$.
The fundamental properties of these curvature-dependent trigonometric
functions are
$$
  \Cos_{\k}^2(x)+ \k \Sin_{\k}^2(x) =1 \,,
$$
and
\begin{eqnarray}
  \Cos_{\k}(2x)=\Cos_{\k}^2(x) - \k \Sin_{\k}^2(x)\,,& \qquad &
  \frac{d}{dx}\Sin_{\k}(x)=\Cos_{\k}(x)\,, \cr
  \Sin_{\k}(2x)=2\Sin_{\k}(x)\Cos_{\k}(x)\,, & \qquad &
  \frac{d}{dx}\Cos_{\k}(x)=-\k \Sin_{\k}(x)\,.
\label{ktrig}\end{eqnarray}

    It is well known that the appropriate coordinates for the study of
central potentials are polar $(r,\phi)$ coordinates (see Appendix);
according to the intrinsic viewpoint, the coordinate $r$ in any space of
constant curvature $\k$ is the actual distance measured along the geodesics
emanating from an arbitrarily chosen origin point $O$, while $\phi$ has
the same meaning as in $E^2$, the angle between this geodesic and a
reference geodesic through $O$ \cite{Kl78}. The range of $r$ depends on
$\k$; specifically $r\in[0, \pi/2\sqrt{\k}]$ for $\k>0$ and
$r\in[0, \infty]$ for $\k\leq0$.
The range of $\phi$ is the interval $[0, 2\pi]$.  So we start with the
following $\k$-dependent expression
$$
  ds^2 = d\,r^2 + \Sin_\k^2(r)\,d{\phi}^2 \,,
$$
that represents the differential line element on the spaces
$(S^2,\IE^2,H^2)$ with constant curvature $\k$. This metric reduces to
$$
  ds_1^2 =    d\,r^2 + (\sin^2 r)\,d{\phi}^2 \,,{\quad}
  ds_0^2 =    d\,r^2 + r^2\,d{\phi}^2 \,,{\quad}
  ds_{-1}^2 = d\,r^2 + (\sinh^2 r)\,d{\phi}^2\,,
$$
in the three particular cases of the unit sphere $\k=1$, Euclidean 
plane $\k=0$,
and `unit' Lobachewski plane $\k=-1$.

The three $\k$-dependent vector fields $Y_1$, $Y_2$, $Y_J$, with coordinate
expressions given by
\begin{eqnarray}
     Y_{1}(\k) &=& (\cos{\phi})\,\fracpd{}{r} -
\Bigl(\frac{\Cos_{\k}(r)}{\Sin_{\k}(r)}\sin{\phi}\Bigr)\,\fracpd{}{\phi} \,,\cr
     Y_{2}(\k) &=& (\sin{\phi})\,\fracpd{}{r} +
\Bigl(\frac{\Cos_{\k}(r)}{\Sin_{\k}(r)}\cos{\phi}\Bigr)\,\fracpd{}{\phi} \,,\cr
     Y_J   &=& \fracpd{}{\phi} \,,    {\nonumber}
\end{eqnarray}
are generators of three different one-parameter groups of
diffeomorphisms preserving the metric $ds^2$ (isometries of the
Riemannian manifold) \cite{RaS03}. In fact, the commutators of
these vector fields are given by
$$
  [Y_1(\k)\,,Y_2(\k)] =  -\,{\k}\,Y_J \,,{\quad}
  [Y_1(\k)\,,Y_J] =  -\,Y_2(\k)       \,,{\quad}
  [Y_2(\k)\,,Y_J] =     Y_1(\k)    \,,
$$
so that they close the Lie algebra of the group of isometries of the
spherical (Euclidean, hyperbolic) space.
Notice that only when $\k=0$ (Euclidean plane) $Y_{1}$ and $Y_{2}$ 
will commute.
Moreover, the Lagrangian for the geodesic (free) motion is given by the
kinetic term arising from the Riemannian metric; that is
$$
  L_0(r, \phi, v_r, v_\phi; \k) = T(\k) =
  (\frac{1}{2})\,\Bigl(\,v_r^2 + \Sin_\k^2(r) v_\phi^2\,\Bigr) \,,
$$
and is invariant under the actions of $Y_1(\k)$, $Y_2(\k)$, and $Y_J(\k)$.

A general standard Lagrangian ($\k$-dependent kinetic term minus a
potential) has the following form
$$
  L(r, \phi, v_r, v_\phi; \k) = (\frac{1}{2})\,\Bigl(\,v_r^2 + \Sin_\k^2(r)
v_{\phi}^2\,\Bigr) -  U(r,\phi;\k)  \,,
$$
in such a way that for $\k=0$ we recover a standard Euclidean system
$$
  \lim_{\k\to 0}\,L = (\frac{1}{2})\,(\,v_r^2 + r^2\,v_{\phi}^2\,)
  - V(r,\phi)
  \,,{\quad}   V(r,\phi) = U(r,\phi;0) \,.
$$
In some particular cases this Lagrangian system possesses the
vector fields $Y_1$, $Y_2$, or $Y_J$, as exact Noether symmetries.
If we denote by $Y_s^t$, $s=1,2,J$, the natural lift to the tangent bundle
(phase space) of the vector field $Y_s$  and by
$\te_L$ the Cartan semibasic one-form \cite{MaS85}
$$
  \te_L = \frac{\partial L}{\partial v_r}\, dr + \frac{\partial L}{\partial
  v_\phi}\, d\phi  =  v_r\,dr + \Sin_\k^2(r) v_{\phi}\,d\phi \,,
$$
then the cases with exact Noether symmetries are the following:

\begin{enumerate}
\item  If the potential $U$ is of the form $U = U(z_2)$, with
$z_2 = \Sin_{\k}(r)\sin{\phi}$, then
$$
  P_1(\k) = i(Y_1^t(\k))\,\theta_L =
  (\cos{\phi})\,v_r - (\Cos_\k(r) \Sin_\k(r)\sin{\phi})\,v_{\phi}
$$
is a constant of motion.
\item  If the potential $U$ is of the form $U = U(z_1)$,
$z_1 = \Sin_{\k}(r)\cos{\phi}$, then
$$
  P_2(\k) = i(Y_2^t(\k))\,\theta_L =
  (\sin{\phi})\,v_r + (\Cos_\k(r) \Sin_\k(r)\cos{\phi})\,v_{\phi}
$$
is a constant of motion.
\item If the potential $U$ depends only on the coordinate $r$
(central potential), then
$$
  J(\k) = i(Y_J^t)\,\theta_L = \Sin_{\k}^2(r)\,v_{\phi}
$$
is a constant of motion.
\end{enumerate}
We remark that the dependence on the curvature $\k$ is only explicit
in the radial dependence of these functions; the angular dependence,
contained in $\cos{\p}$ or $\sin{\p}$, is $\k$--independent.
As an example, the vector field $Y_J$ is $\k$-independent but the
integral of motion $J(\k)$ is $\k$-dependent.
These quantities $P_1$, $P_2$, $J$, are the $\k$-dependent versions of
the two components of the linear momentum and the angular momentum.

\section{Central potentials on $(S^2,\IE^2,H^2)$}

Making use of the three $\k$-dependent functions, $P_1$, $P_2$ and $J$,
the kinetic energy $T(\k)$ can be rewritten as follows
$$
  T(\k) = (\frac{1}{2})\,\Bigl(\,
        P_1^2(\k) + P_2^2(\k) + \k\,J^2(\k)\,\Bigr) \,,
$$
so that the total energy becomes
$$
  E(\k) = (\frac{1}{2})\,\Bigl(\,
        P_1^2(\k) + P_2^2(\k) + \k J^2(\k)\,\Bigr) + U(r,\phi;\k) \,,
$$
showing that, on spaces of (constant) non-zero curvature, the angular momentum
has a contribution to the total energy of the system, proportional to the
curvature $\k$.

A $\k$-dependent potential $U$ whose expression in `polar'
coordinates $(r,{\phi})$ has the structure
$$
  U = F(r) + \frac{G(\phi)}{\Sin_\k^2(r)}\,,
$$
is Hamilton-Jacobi separable \cite{RaS03}. It is, therefore, integrable
and has the following two quadratic integrals of motion
\begin{eqnarray}
   I_1(\k) &=& P_1^2(\k) + P_2^2(\k) + 2 F(r) +
  \frac{2\,G(\phi)}{\Tan_\k^2(r)} \,,\cr
   I_2(\k) &=& J^2(\k) + 2\,G(\phi)  \,.  {\nonumber}
\end{eqnarray}
Then, in this separable case, the total energy splits as a sum of two
independent constants of motion
$$
  E(\k) = (\frac{1}{2})\,I_1(\k) + (\frac{1}{2})\,\k\,I_2(\k)  \,.
$$
Of course, if $G=0$ then $U=F(r)$ is a central potential and the function
$I_2(\k)$ just reduces to $I_2=J^2(\k)$.

The Lagrangian of a $\k$-dependent central potential is
given by
$$
  L(\k) = (\frac{1}{2})\,\Bigl(\,v_r^2 + \Sin_\k^2(r) v_{\phi}^2\,\Bigr)
        -  U(r;\k) \,,
$$
so that the dynamics is represented by the vector field
$$
  X_L = v_r\,\fracpd{}{r} + v_\phi\,\fracpd{}{\phi}
      + f_r\,\fracpd{}{v_r} + f_{\phi}\,\fracpd{}{v_\phi}
$$
with the functions $f_r$ and $f_{\phi}$ given by
\begin{eqnarray}
   &&f_r = \Sin_\k(r)\Cos_\k(r)v_{\phi}^2 - U'_r \,, \cr
   &&f_{\phi} = -\,2 \bigl(\frac{\Cos_\k(r)}{\Sin_\k(r)}\bigr)\,v_rv_{\phi} \,.
{\nonumber}
\end{eqnarray}
The associated two equations are
\begin{eqnarray}
   &&\frac{d}{dt}\,v_r = \Sin_\k(r)\Cos_\k(r)v_{\phi}^2
    - U'_r  \,,{\qquad} U'_r = \frac{dU}{dr} \,,  \cr
   &&\frac{d}{dt}\,J(\k) = 0 \,,{\qquad} J(\k) = \Sin_\k^2(r) v_{\phi}
\,, {\nonumber}
\end{eqnarray}
where the time-derivative $d/dt$ can be interpreted, in geometric terms,
as the Lie derivative along $X_L$.
The $\phi$--equation just gives the conservation law of the
$\k$-dependent angular momentum (or the law of areas);
concerning the radial equation it can be rewritten as follows
$$
  \dot{v}_r=  \ddot{r} =
  -\,\frac{d}{dt}\biggl[\,(\frac{1}{2})\,\frac{J^2}{\Sin_\k^2(r)}
  + U(r;\k)\,\biggr] \,,
$$
so that, after multiplying by $v_r=\dot{r}$, it leads to
$$
  \frac{d}{dt}\biggl[\,(\frac{1}{2})\,v_r^2 +
  (\frac{1}{2})\,\frac{J^2}{\Sin_\k^2(r)}
  + U(r;\k)\,\biggr] = 0 \,,
$$
that represents the conservation law of the energy
$$
  (\frac{1}{2})\,v_r^2 + (\frac{1}{2})\,\frac{J^2}{\Sin_\k^2(r)} +
  U(r;\k) = E  \,,{\qquad}  \frac{d}{dt}\,E=0 \,.
$$
Solving for $\dot{r}$, we obtain
$$
  \frac{dr}{dt} = \sqrt{ 2\,
  \Biggl[\, E - U(r;\k) - (\frac{1}{2})\,\frac{J^2}{\Sin_\k^2(r)} \,\Biggr]} \,,
$$
that can be solved for $dt$ and integrated
$$
  t = \int \frac{dr}{\sqrt{ 2\,
\bigl[ E -  U(r;\k) - (\frac{1}{2})\,\bigl(J^2/\Sin_\k^2(r)\bigr)
\bigr]}} \,,
$$
so that in the particular $\k=0$ case we recover the integral of
the Euclidean  case appearing in the books of theoretical mechanics
$$
  t = \int \frac{dr}{\sqrt{ 2\,
\bigl[ E - V(r) - (\frac{1}{2})\,\bigl(J^2/r^2\bigr) \bigr] }}.
$$

  Coming back to the general $\k\ne 0$ case, if we consider a change of
variable from $r$ to a new variable $u_\k = \Cos_\k(r)/\Sin_\k(r)$, by using
(\ref{ktrig}) we get
$$
  dr =  -\,\frac{du_\k}{u_\k^2 + \k}  \,,
$$ and by considering the potential
$U=U(r;\k)$ as a function of the new radial variable, $U = U(u_\k;\k)$,
the above integral for $t$ becomes
$$
  t = \int \frac{du_\k}{(u_\k^2 + \k)\,\sqrt{ 2\,
\bigl[ E_P - U(u_\k;\k) - (\frac{1}{2}) J^2 u_\k^2 \bigr) \bigr]}}
\,,
$$
where $E_P$ denotes a constant of motion which which plays an important
role:
$$
  E_P = E - (\frac{1}{2})\,\k J^2  \,.
$$
(Notice the true energy for curvature $\k$ is $E$ and not $E_P$; however $E_P$
can be seen as a kind of $\k$-deformation of the Euclidean energy because
$E_P=E$ for $\k=0$). This integral gives the value of
$t$ as a function of
$u_\k$ and consequently also of $r$;
it can be inverted, at least formally, and we can obtain
$u_\k$ (or $r$) as a function of $t$ and the two constants
$E_P$ and $J$.
The expression of $\phi$ as a function of $t$ is just given by
$$
  \phi = J \int (u_\k^2 + \k)\,dt + \phi_0 \,.
$$
We close this section by calling the attention to the factor
$(u_\k^2 + \k)$ in the above integral for $t$ that clearly resembles
to the factor appearing in the elliptic integrals of the third kind.

\section{The Kepler problem on $(S^2,\IE^2,H^2)$}

  The following spherical (hyperbolic) Lagrangian with curvature $\k$,
$$
  L_K(\k) = (\frac{1}{2})\,\Bigl(\,v_r^2 + \Sin_\k^2(r) v_{\phi}^2\,\Bigr)
         -  U_K(r;\k) \,,{\quad}
  U_K = -\,\frac{k}{\Tan_\k(r)} \,,
$$
represents the $\k$-dependent version of the Euclidean Kepler problem;
the potential $U_K$ reduces to
$$
  U_1(r) =  -\,\frac{k}{\tan r}  \,,{\quad}
  U_0(r) = V(r) = -\,\frac{k}{r}  \,,{\quad}
  U_{-1}(r) = -\,\frac{k}{\tanh r} \,,
$$
in the three particular cases of the unit sphere ($\k=1$),
Euclidean plane ($\k=0$), and `unit' Lobachewski plane ($\k=-1$);
the Euclidean function $V(r)$ appears in this formalism as making
separation between two different behaviours (see Fig. 1). The
global sign has been chosen so that $k>0$ corresponds to an
attractive potential.

  This potential is actually worthy of the name `Kepler in curvature
$\k$' due to two reasons. First, this potential is the spherically
symmetric potential satisfying Gauss law in a {\it
three-dimensional space of constant curvature $\k$}, where the
area of a sphere of radius $r$ is $4\pi \Sin_\k^2(r)$, whence the
flux of the corresponding radial force field across a sphere of
radius $r$, which clearly equals $4\pi\Sin_\k^2(r) (dU_K/dr)$
should be a constant independent of $r$; this condition leads
directly to the potential $U_K$ (As in the Euclidean case, the
potential $-\,(k/\Tan_\k(r))$ does not satisfy the Gauss law in
two dimensions). The second reason is based on superintegrability:
what singularizes $U_K$ among other central potentials with the
correct $-k/r$ Euclidean limit is the property of being  a
superintegrable system for all the values of the curvature $\k$
\cite{RaS99}. In fact, $U_K$ is endowed with the following two
additional integrals of mot\-ion
\begin{eqnarray}
  I_3(\k) &=& P_2(\k)J(\k)  -  k\,\cos{\phi} \,,\cr
  I_4(\k) &=& P_1(\k)J(\k)  +  k\,\sin{\phi} \,,{\nonumber}
\end{eqnarray}
that represent the two-dimensional curvature versions of the Runge-Lenz
constant of motion, whose existence is a consequence of the additional
separability of $U_K(r;\k)$ in two different systems of $\k$-dependent
'parabolic' coordinates
(this superseparability is not studied in this article).
Of course only three of the four integrals are functionally independent;
so we have several possibilities for the choice of a fundamental
set of $\k$-dependent integrals of motion as, for example,
$\bigl\{ E(\k),J(\k),I_3(\k) \bigr\}$,
$\bigl\{ E(\k),J(\k),I_4(\k) \bigr\}$, or
$\bigl\{ E(\k),I_3(\k),I_4(\k) \bigr\}$.

\subsection{The classification of orbits and the equivalent one--dimensional
problem}

  Let us start for an arbitrary central potential $U(r)$.
We have previously obtained
$$
  (\frac{1}{2})\,v_r^2 + U(r;\k) +
  (\frac{1}{2})\,\frac{J^2}{\Sin_\k^2(r)} = E  \,,
$$
so, if we introduce the `effective one-dimensional equivalent $\k$-dependent
potential  $W$' defined as
$$
  W_\k(r) = U(r;\k) + \frac{J^2}{2 \Sin_\k^2(r)} \,,
$$
where the term $J^2/(2 \Sin_\k^2(r))$ plays clearly the role of
the `centrifugal barrier potential', then the above property
reduces to conservation of the energy for the fictitious
$\k$-dependent one--dimensional problem arising from the potential
$W_\k$
$$
  (\frac{1}{2})\,v_r^2 + W_\k(r) = E
$$
so that we can use, also in this $\k$-dependent case, the method of
the classification of the orbits by analyzing the behaviour
of $W_\k$.

Next we will study the main characteristics of the orbits for the
Kepler potential in curvature $\k$ where  $W_\k$ is given by
$$
  W_\k(r) = -\,\frac{k}{\Tan_\k(r)} + \frac{J^2}{\,2\Sin_\k^2(r)} \,,
$$
so it reduces to
\begin{eqnarray}
   &&W_1 =  -\,\frac{k}{\tan r} + \frac{J^2}{\,2\sin^2 r} \,,\cr
   &&W_0  =  -\,\frac{k}{r}  +  \frac{J^2}{\,2 r^2} \,,\cr
   &&W_{-1} = -\,\frac{k}{\tanh r} + \frac{J^2}{\,2\sinh^2 r}\,,
{\nonumber}
\end{eqnarray}
in the three particular one--dimensional problems associated to the
unit sphere ($\k=1$), Euclidean plane ($\k=0$), and `unit'
Lobachewski plane ($\k=-1$).

\noindent{(1)}
Analysis of the potentials $W_1$ and $W_c$ ($c=\sqrt{\k}>0$).

The function $W_1$ satisfies the following limits in the boundaries
$$
  \lim_{r\to 0} W_1(r) = + \infty \,,\quad
  \lim_{r\to \pi} W_1(r) = + \infty \,,
$$
it cuts the $r$-axis at the points $r_{1,2}$ solutions of the
equation $\sin(2r_{1,2})=J^2/k$, and it has a minimum at the point
$r_m^s$ given by $r_m^s = \tg^{-1}(J^2/k)$. It represents,
therefore, a potential well with  barriers of infinite height at
both extremes, $r=0$ and $r=\pi$, and one single minimum placed
inside the left half-interval ($r_m^s<\pi/2$) with a value
$W_m^s=(1/2)(J^2-k^2/J^2)$ (Fig. 2). Thus, all the trajectories
are bounded and all of them describe non-linear one--dimensional
oscillations between the two turning points.

   In the general case of the function $W_c$, the points $r_{1,2}$
are given by the two roots of $\sin(2c r_{1,2})=c(J^2/k)$ and
the minimum is placed in the point $r_m^s = (1/c)\tg^{-1}(c J^2/k)$
with a value given by $W_m^s=(1/2)(c^2J^2-k^2/J^2)$.
All these expressions are $c$--dependent and have the correct limits for
$W_0$:
$$
  \lim_{c\to 0}r_1 = \frac{\,J^2}{2k}  \,,\quad
  \lim_{c\to 0}r_2 = +\infty  \,,\quad
  \lim_{c\to 0}r_m^s = \frac{J^2}{k}  \,,\quad
  \lim_{c\to 0}W_m^s = -\,\frac{k^2}{2 J^2} \,.
$$
(see Fig. 3).

\noindent{(2)}
Analysis of the potentials $W_{-1}$ and $W_{-c}$ ($c=\sqrt{-\k}>0$).

The function $W_{-1}$ satisfies the two following limits
$$
  \lim_{r\to 0} W_{-1}(r) = + \infty \,,\quad
  \lim_{r\to \infty} W_{-1}(r) = -\,k \,,
$$
it cuts the $r$-axis in the point $r_1$ unique solution of the equation
$\sinh(2r_1)=J^2/k$ and, in the case that $k$ and $J$ satisfy the condition
$J^2/k<1$, then it has a unique minimum in the point
$r_m^h$ given by $r_m^h = \tanh^{-1}(J^2/k)$ with the value
$W_m^h=-\,(1/2)(J^2+k^2/J^2)$.
There exist therefore two possible situations:
\begin{enumerate}
\item{}  If $J^2/k<1$ the motion is bounded (periodic) for small
energies $W_m^h{\le}E<-k$, and unbounded for higher energies
$E{\ge}-k$.
\item{}  If $J^2/k{\ge}1$ the function $W_{-1}$ will take the form of
a potential barrier with infinite height at the the origin $r=0$;
only energies satisfying $E{\ge}-k$ will be allowed and all the
trajectories will be unbounded (scattering) open curves.
\end{enumerate}
These two possible behaviours are represented in Figure 4.

   The general hyperbolic function $W_{-c}$ satisfies
$$
  \lim_{r\to 0} W_{-c}(r) = + \infty \,,\quad
  \lim_{r\to \infty} W_{-c}(r) = -\,c k \,,
$$
it cuts the $r$-axis in the point $r_1$ solution of
$\sinh(2cr_1)=c(J^2/k)$ and, if the condition $c(J^2/k)<1$
is satisfied, then it has a unique minimum in the point
$r_m^h = (1/c)\tanh^{-1}c(J^2/k)$ with a value $W_m^h$ given by
$W_m^h=-\,(1/2)(c^2J^2+k^2/J^2)$.
It is clear that the smaller the value of $c$ is, the easier of
satisfying is the condition for the existence of a well,
so that the behaviour (1) is becoming more and more dominant.
Finally, the Euclidean limit is given by
$$
  \lim_{c\to 0}r_1 = \frac{\,J^2}{2k}  \,,\quad
  \lim_{c\to 0}r_m^h = \frac{J^2}{k}  \,,\quad
  \lim_{c\to 0}W_m^h = -\,\frac{k^2}{2 J^2} \,.
$$
The convergence of $W_{-c}$ into $W_0$ is represented in Figure 5.

\subsection{Determination of the orbits of the $\k$-dependent Kepler
problem }

\subsubsection{Method I: Direct Integration }

We have previously obtained, making use of the conservation of the
total energy $E$ and the angular momentum $J$, two
expressions for $\dot{r}$ and $\dot{\phi}$, which can be written as
$$
  dt =  \frac{dr}{\sqrt{ 2\,
  \bigl[ E - U - (1/2)\,\bigl(J^2/\Sin_\k^2(r)\bigr) \bigr] }}, \qquad
  dt = \Bigl(\frac{\Sin_\k^2(r)}{J}\Bigr) \,d\phi  \,.
$$
Eliminating $t$ between both equations we have
$$
  d\phi =  \frac{J\,dr}{\Sin_\k^2(r) \sqrt{ 2\,
  \bigl[ E - U - (1/2)\,\bigl(J^2/\Sin_\k^2(r)\bigr) \bigr]}}  \,,
$$
that after the change of variable $r \to u_\k$ with
$u_\k = 1/\Tan_\k(r)$,  $dr = - \Sin_\k^2(r)\,du_\k$, becomes
$$
  d \phi = -\,\frac{du_\k}{\sqrt{ R(u_\k)}} \,,
$$
where$R(u_k)$ denotes the following function
\begin{eqnarray}
  R(u_k) &=& \frac{2 E}{J^2} - \frac{2 U}{J^2} - \bigl(u_\k^2 + r_\k^2\bigr) \cr
         &=& \frac{2 E_P}{J^2} - \frac{2 U}{J^2} - u_\k^2 \,.
{\nonumber}
\end{eqnarray}
All this is valid for a general potential.
Next we particularize for the $\k$-Kepler problem,
$U_K(r;\k) = -\,{k}/{\,\Tan_\k(r)}$, and then we obtain
$$
  \phi = \phi_0 -\,\frac{du_\k}{\sqrt{ R(u_\k) }}  \,,\qquad
  R(u_k) = \al + \be u_\k + \ga u_\k^2  \,,
$$
with coefficients $\al$, $\be$ and $\ga$ given by
$$
  \al=\frac{2 E_P}{J^2} \,,{\quad} \be=\frac{2 k}{J^2} \,,{\quad} \ga=-1 \,.
$$
In this particular case the integration is elementary and we arrive at
$$
  \phi = \phi_0 -
  \cos^{-1}\Bigl( \frac{u_\k - (k/J^2)}{(k/J^2)\sqrt{1+z_\k}} \Bigr)
  \,,{\quad} z_\k= \Bigl(\frac{2  J^2}{k^2}\Bigr) E_P \,,
$$
leading to
\begin{equation}
  u_\k(\phi) = \Bigl(\frac{k}{J^2}\Bigr) \bigl[1 + e_\k \cos(\phi-\phi_0)\bigr]
  \,,{\quad} e_\k=\sqrt{1+z_\k}\label{KEqOrbit} \,.
\end{equation}

This is the polar equation of the orbit in either space
$(S^2_\k,\IE^2,H^2_\k)$, for any value of the curvature $\k$.
It reduces to
\begin{eqnarray}
  &&\frac{1}{\tan r} = \Bigl(\frac{k}{J^2}\Bigr)
            \bigl[1 + e_1 \cos(\phi-\phi_0)\bigr] \,,\cr
  &&\frac{1}{r}  =  \Bigl(\frac{k}{J^2}\Bigr)
            \bigl[1 + e_0 \cos(\phi-\phi_0)\bigr] \,,\cr
  &&\frac{1}{\tanh r} = \Bigl(\frac{k}{J^2}\Bigr)
            \bigl[1 + e_{-1}\cos(\phi-\phi_0)\bigr]\,,{\nonumber}
\end{eqnarray}
in the three particular cases of the unit sphere ($\k=1$),
Euclidean plane ($\k=0$), and `unit` Lobachewski plane ($\k=-1$)
respectively.

  We recall that, as only three of the several integrals of motion can
be functionally independent, there must exist some relations between them.
Concerning the couple $I_3(\k)$, $I_4(\k)$, that, as stated above, represent
the $\k$-dependent version of the Euclidean Runge-Lenz constants of motion,
they are related with $E_P$ and $J$ by
$$
  I_3^2 + I_4^2 = 2 E_P J^2 + k^2
$$
so that we arrive to
$$
  e_\k = \frac{1}{k}\,\sqrt{I_3^2 + I_4^2}
$$
that represents the natural extension to the curvature-dependent case
of a well known property of the Euclidean case.

  Let us close this direct integration approach with three observations.
First, it turns out that in any of the three manifolds
$(S^2_\k,\IE^2,H^2_\k)$, and for any value of $\k$, this curve is
a {\it conic with a focus at the origin}, where `conic' has to be
taken in a metric sense,  relative to the intrinsic metric in each
space. This follows from the geometrical study to be presented in
Section 5. Second,  notice that the quantity $e_\k$ is related to
the `partial energy' $E_P$ exactly as the Euclidean eccentricity
of the conic is related to the total energy, yet only in the
Euclidean case ($\k=0$) this function $E_P$ coincides with $E$. In
the general case ($\k\ne 0$) the characteristics of the orbit will
be more easily associated to the values of the $e_\k$ and/or $E_P$
than to the total energy $E$. Third, both the method and the
results obtained show a close similarity with the Euclidean ones.
In fact, the important point is that the classical and well known
change of variable $r{\to}u=1/r$ admits as a generalization the
$\k$--dependent change $r{\to}u_\k = 1/\Tan_\k(r)$ which affords a
significant simplification for all values of $\k$; conversely,
this $\k$--dependent change reduces to the Euclidean change
$r{\to}u_0=1/r$ for $\k=0$.

\subsubsection{Method II: Equation of Binet }

The definition of the angular momentum $J$ determines a
$\k$-dependent relation between the differentials of the
time $t$ and the angle $\phi$
$$
  J\,dt = \Sin_\k^2(r)\,d\phi \,.
$$
The corresponding relation between the derivatives with respect to
$t$ and $\phi$  is
$$
  \frac{d}{dt} = \Bigl(\frac{J}{\Sin_\k^2(r)}\Bigr)\frac{d}{d\phi} \,,
$$
so that the second derivative with respect to $t$ is given by
$$
  \frac{d^2}{dt^2} = \Bigl(\frac{J}{\Sin_\k^2(r)}\Bigr)\frac{d}{d\phi}
  \Bigl[\Bigl(\frac{J}{\Sin_\k^2(r)}\Bigr)\frac{d}{d\phi}\Bigr] \,.
$$
Introducing this notation in the radial equation, it becomes
$$
  \Bigl(\frac{J}{\Sin_\k^2(r)}\Bigr)\frac{d}{d\phi}
  \Bigl[\Bigl(\frac{J}{\Sin_\k^2(r)}\Bigr)\frac{dr}{d\phi}\Bigr]
  - \Bigl(\frac{\Cos_\k(r)}{\Sin_\k^3(r)}\Bigr) J^2
  = - U'_r \,.
$$
This equation can be simplified in two steps: firstly, the
lefthand side can be rewritten by making use of
$$
  \frac{d}{d\phi} \Bigl(\frac{\Cos_\k(r)}{\Sin_\k(r)}\Bigr) =
   -\,\Bigl(\frac{1}{\Sin_\k^2(r)}\Bigr) \frac{dr}{d\phi} \,;
$$
secondly, we introduce the change $r\to u_\k$
in such a way that the potential $U= U(r;\k)$ be considered as a
function of $u_\k$;
then we have
$$
  U'_r = -\,\Bigl(\frac{1}{\Sin_\k^2(r)}\Bigr)\,U'_u \,,{\qquad}
  U'_u = \frac{dU}{du_\k}  \,.
$$
In this way we arrive at the differential equation of the orbit
$$
  \frac{d^2u_\k}{d\phi^2} + u_\k = -\,\Bigl(\frac{1}{J^2}\Bigr) U'_u \,,
$$
that permits us to obtain $\phi$ as a function of $u_\k$ for the 
given potential
$U(r;\k)$ when considered as function of $u_\k$:
$$
  \phi = \int \Bigl\{ c -\,\Bigl(\frac{2}{J^2}\Bigr) U - u_\k^2
  \Bigr\}^{-(1/2)} d u_\k  \,.
$$

   Let us now particularize for the Kepler problem.
In this case the potential $U_K$ is given by $U_K=-\,ku_\k$, and
the equation reduces to a linear equation with constant coefficients
$$
  \frac{d^2u_\k}{d\phi^2} + u_\k = \frac{k}{\,J^2} \,,
$$
which has the general solution
\begin{equation}
   u_\k = A \cos(\phi-\phi_0) + \frac{k}{\,J^2}
   = \Bigl(\frac{k}{J^2}\Bigr) \bigl[1 + e \cos(\phi-\phi_0)\bigr] \,,
\label{KEqOrbit2}
\end{equation}
where $A$ (or $e=A\,(J^2/k)$) and $\phi_0$ are the two constants of
integration.

Remark that the differential equation of the orbit, usually known
as the Binet's Equation, is preserved by the $\k$--deformation.
That is, $u_0$ deforms to $u_\k$ but the the equation by itself
remains invariant.

\subsection{Analysis of the orbits}

 From the geometrical viewpoint, one of the integration constants in
(\ref{KEqOrbit}) or (\ref{KEqOrbit2}) can be made to disappear by simply
choosing $\phi$ to be measured from the orbit position closest to the focus.
Thus the orbit is
\begin{equation}
  \Tan_\k(r) = \frac{D}{1+\e \cos\phi}\label{conicEq0}
\end{equation}
which depends on two geometric parameters, the constants $D$ and $e$.
Comparing with (\ref{KEqOrbit}) or (\ref{KEqOrbit2}), these two geometric
parameters are related to the angular momentum $J$ and the total energy
$E$ by means of the expressions
\begin{equation}
  D = \frac{J^2}{k},   \qquad
  \e_\k = \sqrt{1+\frac{2J^2}{k^2}\Big(E-\k\frac{J^2}{2}\Big)} \,,
\label{AeVersusJE} \end{equation}
remaining valid  for any value of the curvature $\k$ and reducing to
the known expressions for the Euclidean case, $\k=0$.
Thus, as advanced before, the relation between the  geometric constants
$D$, $\e$, and the physical constants $J$, $E-(1/2)\,\k J^2=E_P$, is the
same for all values of curvature; here we see again the quantity $E_P$ to
play an role: it can be considered as a kind of `translational' part of the
energy, not taking into account the contribution to the energy of the angular
momentum.

The nature of the orbit depends on the values of these constants.
We discuss separately the cases with positive and negative $\k$.

\noindent (i)
Spherical space, $\k>0$. The polar equation of the orbits is:
$$
  \frac{1}{\Tan_\k(r)}
  = \frac{\sqrt{\k}}{\tan{\sqrt{\k}\,r}}  = \Bigl(\frac{k}{J^2}\Bigr)
  \bigl[1 + e_\k \cos(\phi-\phi_0)\bigr]\,.
$$
The orbit is always a closed curve and for any value of $\e_\k$ this curve
is an spherical ellipse with a focus at the origin. For a fixed $J$ the
minimal value of the total energy corresponds to orbits with $\e_\k=0$,
which are circles with radius $r=J^2/k$ and total energy
$E_{\rm cir}=(1/2)(- k^2/J^2 + \k J^2)$
(which may be either negative or positive).
The possible energies for given $J$ fill the interval $[E_{\rm cir},\infty]$
and correspond to the parameter $\e_\k$ in the interval $[0, \infty]$
and to $E_P$ in the interval $[-(1/2)(k^2/J^2),\infty]$.
In the following description the focus will be conventionally placed at
the North pole, and upper and lower half-spheres refer to the North and
South hemispheres.
This orbit has always a vertex closest to the focus given by the unique
solution of the equation $1/\Tan_\k(r)=(k/J^2)(1 + e_\k)$, and depending on
the value of $\e_\k$ we may however distinguish three possible behaviours for
the second vertex:
\begin{itemize}
\item  When $e_\k<1$ the orbit is completely contained in the
upper half-sphere centered at the focus, and the values of $r$
always remain less than the length $\pi/(2\sqrt{\k})$ of a
quadrant on the sphere of curvature $\k$. In some respects this
reminds the case of Euclidean ellipses.

\item  If $e_\k=1$, the ellipse has the second  vertex at $r=\pi/(2\sqrt{\k})$
(reducing to $r=\pi/2$ when $\k=1$), and is a curve touching tangentially the
equator associated to the focus at the origin. When $\k\to0$, the limit of the
distance $r=\pi/(2\sqrt{\k})$ between the origin and the equator on a sphere
$S^2_\k$ is $r=\infty$ and thus this curve is analogous to a 
Euclidean parabola,
touching the spatial infinity at a point.

\item  When $e_\k>1$, the orbit has the second vertex at $r>\pi/(2\sqrt{\k})$;
it is a big spherical ellipse crossing the equator of the focus and entering
into the lower half-sphere.
This is somehow analogous to an Euclidean hyperbola.
\end{itemize}

\noindent (ii)
Hyperbolic space $\k<0$.
Let us consider the general case of arbitrary negative curvature. The
orbit equation is now
$$
  \frac{1}{\Tan_\k(r)} =
  \frac{\sqrt{-\k}}{\tanh\,{\sqrt{-\k}\,r}} = \Bigl(\frac{k}{J^2}\Bigr)
        \bigl[1 + e_{\k} \cos(\phi-\phi_0)\bigr] \,,\quad  \k<0  \,,
$$
and the relations (\ref{AeVersusJE}) also apply but now with $\k<0$.
For a fixed value of $J$, the diagram of the effective equivalent potential
reveals  two essentially different situations for the behaviour of $W(r)$,
corresponding to values of $J$ smaller or larger than the  particular
value $J_\infty$ of $J$:
$$
  J_\infty^2 = \frac{k}{\sqrt{-\k}}  \,.
$$

  When $J<J_\infty$ ($J^2/k < 1/\sqrt{-\k}$), the equivalent potential is
qualitatively as its Euclidean counterpart; it has a minimum at $r=r_{\rm cir}$
determined by $\Tan_\k(r_{\rm cir})=J^2/k$, with a {\it negative} value for
$W(r_{\rm cir})=E_{\rm cir}=-{k^2}/{2J^2}+\k J^2/2$ and when $r$ increases
from $r_{\rm cir}$ to $r\to\infty$, the effective potential tends from below to
a {\it negative} constant value $-k\sqrt{-\k}$.
When $J=J_\infty$ then the minimum occurs at $r=\infty$.
When $J>J_\infty$ there is no longer a minimum in the equivalent potential,
which is a monotone function of $r$, tending  when $r\to\infty$ to the value
$-k\sqrt{-\k}$ from above. This discussion alone suffices to establish two
landmark values for the energies (at a fixed value of $J$):
\begin{itemize}
\item  $E=E_{\rm cir}=(1/2)(- k^2/J^2 + \k J^2)$, with $J^2/k<
1/\sqrt{-\k}$, corresponding to the circular orbits with
$E_P{}_{\rm cir}=-(1/2)(k^2/J^2)$, \item
$E=E_{\infty}=-k\sqrt{-\k}$, with $E_P{}_\infty=-k\sqrt{-\k}-
(1/2) \k J_\infty^2=-(1/2)k\sqrt{-\k}$ for orbits where the
particle barely reaches spatial infinity with velocity 0.
\end{itemize}

  To go further in any analysis about the exact nature of the orbits is
easier starting from some geometric knowledge  of the properties of conics,
to be sketched in the next section. However, at this point we can provide
a first classification into closed and open orbits, and we may discuss
its main characteristics.

When the energy is in the interval $[E_{\rm cir}, E_{\infty})$,
the motion is bounded and periodic, while for $E$ in the interval
$(E_{\infty}, \infty]$, the motion is not periodic and the orbit
goes to the infinity; notice that $E=0$ is contained {\it within}
this second interval, and thus is {\it not} the separating value
between bounded and unbounded orbits. The angular momentum
provides a second constant of motion, which may have values in the
two subintervals $J^2/k\in[\Tan_\k(r_{\rm cir}), 1/\sqrt{-\k}]$ or
$J^2/k\in[1/\sqrt{-\k}, \infty]$;  it is pertinent to discuss
these two possibilities successively.

Let us first discuss the subcase $J<J_\infty$, where the
equivalent potential has a minimum.
\begin{enumerate}
\item If the parameter $\e_{\k}=0$ then the solution is clearly a
circular orbit
(circle in the hyperbolic plane) of radius $r_{\rm cir}$ such that
$\Tan_\k(r_{\rm cir})=J^2/k$ which happens precisely at the minimum
for the equivalent potential.
The total energy for a circular orbit is given by
$E_{\rm cir}=-\,(1/2)(k^2/J^2 -\k J^2)$
that corresponds to $E_P{}_{\rm cir} = - (1/2)(k^2/J^2)$.

This is, the angular momentum of any circular orbit
should remain lower than the limit value  $J_\infty$ which
is approached for the circular orbit with  $r\to\infty$.
In the Euclidean case the value of $J$ is not bounded for circular orbits
($J_\infty$ tends to infinity for $\k\to 0$).

\item If $\e_{\k}$ is greater than $0$ but in the interval
$0< \e_{\k} <1-\sqrt{-\k}\,(J^2/k)$ then the orbit  will be a closed curve
not reaching the spatial infinity. As we will check in
Sec.\ 5, this curve is indeed an hyperbolic {\it ellipse}, with a focus at the
potential origin. The interval of allowed values for the total energy $E$ turns
out to be
$$
  -\,\frac{k^2}{2J^2} + \frac{1}{2}\k\,J^2 <\ E\ < -\,k\,\sqrt{-\k} \,,
$$
or equivalently, if we consider the ``translational'' part of the energy:
$$
  -\,\frac{k^2}{2J^2} <\ E_P\ < \frac{1}{2}\k\,J^2 -\,k\,\sqrt{-\k} \,.
$$

\item If $\e_{\k}$ is in the remaining interval,
$1-\sqrt{-\k}\,(J^2/k)<\e_{\k}$, corresponding to the values
$-\sqrt{-\k}\,k<E<\infty$ of the energy $E$, then the the motion
is non periodic and the orbits are unbounded open curves. The
border value for $e$ between these closed periodic and open not
periodic types of orbits is the value
$\e_{\k}=1-\sqrt{-\k}\,(J^2/k)$, which should be thought of as
corresponding to the `last ellipse', or {\it horoellipse}. It is
not immediately clear whether the orbits with $e$ above this value
are {\it parabolas} or {\it hyperbolas} in the hyperbolic plane.
However, there is natural to expect some special status for the
orbits with $E=0$, $\e_{\k}=\sqrt{1 -\k(J^4/k^2)}\,>1$, which we
might expect to be {\it parabolas} in the hyperbolic plane. This
would imply that at least all orbits in a finite interval of
values of the $\e_{\k}$ parameter, whose lower bound is
$1-\sqrt{-\k}\,(J^2/k)$ and which contains
$\e_{\k}=\sqrt{1-\k\,(J^4/k^2)}$ are parabolas; notice this
interval should reduce to the single value $\e_\k=1$ in the
Euclidean limit $\k\to 0$,  so the existence of many different
parabolic orbits does not spoil the known fact that there is only
a parabolic orbit in $E^2$. All these conjectural properties are
indeed true, as we will confirm in Sect.\ 5.
\end{enumerate}

We can summarize these three points as follows:
the different types of trajectories in $H^2_\k$, $\k<0$ as a function of the
parameter $\e_{\k}$ are the following
\begin{eqnarray}
  &&{\rm hyperbolic\ circle} {\hskip 50pt}
      0 =\ \e_{\k}\      \cr
  &&{\rm hyperbolic\ ellipses}{\hskip 40pt}
      0 <\ \e_{\k}\ <\  1-\sqrt{-\k}\,(J^2/k)  \cr
  &&{\rm hyperbolic\ horoellipse}{\hskip 22pt}
      1-\sqrt{-\k}\,(J^2/k)=\e_{\k}    \cr
  &&{\rm hyperbolic\  open\  conics}{\hskip 18pt}
      1-\sqrt{-\k}\,(J^2/k)<\ \e_{\k}\ <\infty \,.
{\nonumber}\end{eqnarray}
Alternatively, the landmark values can also be given for the 
classification as a
function of the total Energy $E$:
\begin{eqnarray}
  &&{\rm hyperbolic\ circle} {\hskip 50pt}
     E=-\,(1/2)(k^2/J^2 -\k J^2)     \cr
  &&{\rm hyperbolic\ ellipses}{\hskip 40pt}
     -\,(1/2)(k^2/J^2 -\k J^2) <\ E\ <\ -\,k\,\sqrt{-\k}  \cr
  &&{\rm hyperbolic\ horoellipse}{\hskip 22pt}
     E = -\,k\,\sqrt{-\k}      \cr
  &&{\rm hyperbolic\  open\  conics}{\hskip 18pt}
     -\,k\,\sqrt{-\k} <\ E\  <\infty
{\nonumber}\end{eqnarray}
Thus the behaviour of the hyperbolic dynamics shows some important
differences with respect to the Euclidean case.  We postpone until the
next section the establishment of the particular value of $e_\k$, or $E$,
separating the open conics into parabolic and hyperbolic regimes, and we
simply state that $\e_{\k}=\sqrt{1-\k\,(L^4/k^2)}$, corresponding to $E=0$,
gives an orbit which is a parabola in $H^2$.
The value of $\e_\k$, or the corresponding energy $E$, separating from
parabolic to hyperbolic trajectories is a $\k$-dependent positive value
to be  determined; this will be done in the next section.
When the parameter $e_\k$ is greater than the upper value for parabolic
orbits, then we will have hyperbolas in $H^2$ as orbits.

  The remaining cases happen when $J>J_\infty$.
For these values of $J$ the equivalent potential is a decreasing function
of $r$ with no minimum at all, and hence the possible motions should have
energies above $E_\infty$; these would be always non-periodic motions and
open orbits. We will classify them into parabolas and hyperbolas in the
next section.

\section{Conics on spaces of constant curvature}

In this section we give a geometric description of conics in
the three $\k>0$, $\k=0$, $\k<0$, constant curvature spaces $S^2_\k$,
$E^2$, $H^2_\k$, emphasizing those aspects relevant in relation with
Kepler motion in these spaces.
To stay within a reasonable extension limits we restrict to a
mainly `statement of facts' presentation, which should either confirm
the conjectures advanced in the dynamical part, or serve as a geometric
foundation for them.
The important thing is the perfect matching between the dynamical
approach in Sec.\ 4 and the results here.

As customary in this context, {\it lines} will mean the {\it
geodesics} of the constant curvature space.  The metric intrinsic
{\it geometric} definition of conics, that can be applied to any
2-d space of constant curvature $\k$ involves {\it focal
elements}, i.e., either oriented points or cooriented lines.
In any such space, and by definition:

  An {\it ellipse/hyperbola} will be the set of points with a constant
{sum/difference} $2a$ of distances $r_1, r_2$, to two fixed points
$F_1, F_2$, called {\it foci} and separated a distance $2f$.

  A {\it parabola} will be the set of points with a constant
{sum/difference} $2\alpha$ of distances $r_1, \wt{r}_2$, to a fixed
point $F_1$, called {\it focus}, and to a fixed cooriented line $f_2$,
called {\it focal line}; the {\it oriented} distance $2\f$
between $F_1$ and $f_2$ plays here the role of focal separation.

  An {\it ultraellipse/ultrahyperbola} will be the set of points with a
constant {sum/difference} $2a$ of oriented distances $\wt{r}_1,\wt{r}_2$,
to two fixed intersecting lines $f_1, f_2$, separated by an angle
$2F$ and called {\it focal lines}.

In the generic case of constant curvature $\k\neq0$ these three
pairs of curves, each pair sharing the same focal elements, are the
{\it generic} conics; Euclidean plane is {\it not generic}
among the family of constant curvature spaces, thus some
Euclidean properties of conics  are very special and do not provide a
good starting viewpoint to discuss the $\k\neq0$ properties.

Further to generic conics, {\it particular distinguished conics}
appear for instance when the focal separation vanishes ($f=0,
\f=0, F=0$) and {\it limiting conics} when some focal elements go
to infinity (if possible at all); both particular and limiting conics
can be obtained as suitable limits from the generic ones.

We first list some particular conics. When the focal separation vanishes $f=0$,
the ellipse is a {\it circle} and the hyperbola is a pair of intersecting
{\it lines}.  No useful intuition about parabolas can be drawn
from the non-generic Euclidean case; a parabola with vanishing focal
separation $\f=0$ is only a particular instance of parabola, to be named
here  {\it equiparabola}, but when $\k\neq0$ there are other parabolas,
with any positive or negative $\f\neq0$, which are not equiparabolas.
The main point is that general parabolas are {\it not} to be considered
here as limiting conics, as usually done in $E^2$.
An ultraellipse with focal angle $F=0$ is a {\it ultracircle}
(a equidistant curve, the set of points equidistant from a fixed line),
while an ultrahyperbola with $F=0$ is a pair of {\it lines}.

The three {\it generic types} of conics, as well as these {\it
particular} conics with zero focal separation do exist in any space
of constant curvature $\k$.
Further to that, when the curvature is {\it negative}, there are new
types of {\it limiting conics}, where either the foci or the focal lines
may go to infinity.

\subsection{Conics in geodesic polar coordinates in a space of curvature $\k$}

Now let us describe conics in the three essentially different spaces
$S^2_\k$, $E^2$, $H^2_\k$,.

In the sphere $S^2_\k$, with $\k>0$ the situation is rather simple.
First, as there are neither points nor lines at the infinity, there are no
limiting cases in the sense they will appear in $H^2_\k$. Second, as
there is a 1-1 correspondence (the usual polarity) between oriented points
and cooriented lines (North Pole versus Equator), with a
constant distance $\pi/2\sqrt{\k}$ separating them, it follows  that any
focal element (either point or line) implies the existence of a polar
focal element (line or point); thus any conic of one of the three given
type pairs may be considered as a conic of any other prescribed type pair.
And further, as the distance between two antipodal points is constant
$\pi/\sqrt{\k}$, by changing a focus
$F_2$ to its antipodal point $\overline{F_2}$, (which according to the
definition is again a focus of the conic), an hyperbola with focus
$F_1, F_2$, will be as well an ellipse with focus $F_1,\overline{F_2}$.
A similar possibility exists for focal lines which may be
changed to their antipodal lines, with coorientation changed.  Thus all
spherical conics can be seen as {\it spherical ellipses}, including the
two particular zero focal separation cases, i.e., circles (spherical parallel
circles) and pairs of lines (a pair of intersecting large circles in the
sphere).
For the purposes of getting a common view we may still think
in three generic types, keeping in mind that all three coincide and
there is complete freedom in understanding  a given spherical conic
as either an ellipse, or a parabola or a hyperbola
(just as a geodesic circle can be seen at the same time as a geodesic circle,
with center at $O$ or as an equidistant curve to the polar of $O$);
more on this will be said later.

In the Euclidean plane $E^2$ ultraellipses and ultrahyperbolas
(with any focal angle $F$) are just pairs of parallel lines, in
directions parallel to the two bisectors of the focal angle; in
the standard algebraic classification of Euclidean conics these appear as
degenerate conics. But when compared to $S^2_\k$ or $H^2_\k$, another
more important degeneracy happens in $E^2$: two families of parabolas with the
same focus and any two parallel focal lines will coincide (because a pair of
parallel lines are equidistant in $E^2$). This means that the {\it parabolic
focal separation} between the focus and the focal line  can be chosen
arbitrarily for a fixed Euclidean parabola, which only determines its
focus and the direction of the focal line. As a consequence parabolas are
no longer generic conics in $E^2$.
While parabolas have always some intermediate status between ellipses
and hyperbolas, in the Euclidean case parabolas appears exclusively as
limiting cases between ellipses and hyperbolas, and not as the full
fledged species of conics they are in the generic $\k\neq0$ case.

The hyperbolic case $H^2_\k$ with {\it negative} constant curvature
is richer. The limiting conics obtained from a ellipse/hyperbola through
a point $P$ when a focus stays fixed and the other goes to infinity are
called {\it horoellipse/horohyperbola}. The limiting conics obtained from a
ultraellipse/ultrahyperbola through a point $P$ when a focal line stays
fixed and the other goes to infinity are called {\it
horoultraellipse/horoultrahyperbola}.
These four types of conics are also limiting forms of parabolas, with
either focus fixed and focal line going to infinity, or focal line fixed
and focus going to infinity. In $E^2$ all these limiting conics collapse
precisely to either Euclidean parabolas or pairs of lines.

  There are more limiting conics in $H^2_\k$: the {\it horocycle}, obtained
either from a circle through $P$ when its center goes to infinity, or from
an ultracircle through $P$ when the baseline goes to infinity, and the
{\it osculating parabola}, the limit of a parabola when the focus go to
infinity along the fixed focal line.

  In order to make contact with the results in the dynamical part, let
us draw our attention precisely on the conics with a proper focus.
 From now on we completely disregard ultraellipses and ultrahyperbolas,
which have two focal lines, and osculating parabolas, which have not a
proper focus.
We consider exclusively ellipses/hyperbolas and parabolas, as well as
their limiting cases ---horoellipses and horohyperbolas--- and particular
cases ---circles, lines and equiparabolas corresponding to vanishing focal
separations---. All these conics have a proper focus, which we
may place at the origin point, as well as another focal element, which
may be another focus or a focal line. In either case the conics
have as a symmetry axis the line joining the two foci or going
through the focus and orthogonal to the focal line.
To match with the familiar Euclidean expressions, we introduce polar
coordinates in $S^2_\k$, $E^2$, $H^2_\k$, taking the origin at the fixed
focus  and the symmetry axis of the conic as the half-line $\phi=0$.
With this choice, the common equation of the whole family of conics
described above may be written in the form:
\begin{equation}
  \Tan_\k(r) = \frac{D}{1+\e \cos\phi}\label{conicEq1}
\end{equation}
with $D$ and $\e$ non-negative. This can be derived by
trigonometric considerations from the definition of the conic in
the space of constant curvature $\k$  using the relations given in
\cite{HeOS00} and will be discussed elsewhere. This relation
justifies the claim made in Sec.\ 5: the orbits of the Kepler
problem in  either $S^2_\k$, $E^2$, $H^2_\k$, are conics with a
focus at the origin of the potential. The dependence on the polar
angle $\phi$ is exactly the same as in the Euclidean case and
this may suggest to consider $\e$ as the non-Euclidean analogue of
the {\it eccentricity of the conic}, though when $\k\neq0$ the
link between the values of $\e$ and the {\it type} of the conic is
not so direct as in $E^2$.
We should mention that there are at least two other quantities
keeping different properties of the Euclidean eccentricity, and
no single quantity keeps all properties; in this paper we will only
be concerned with $\e$, and for brevity we shall refer to it as eccentricity.

For an ellipse/hyperbola with focal distance $2f$ and $2a$ as the
sum/difference of distances to the focus, and for a
parabola with focal separation $2\f$ and $2\alpha$ as sum/difference of
distances to focus and focal line, the eccentricities turn out to be
$$
  \e_{\rm ell/hyp}=\frac{\Sin_\k(2f)}{\Sin_\k(2a)}, \qquad
  \e_{\rm par}=\frac{\Cos_\k(2\f)}{\Cos_\k(2\alpha)}
$$
reducing in the Euclidean $\k=0$ case to the well known $f/a$ for
ellipses/hyperbolas and $1$ for parabolas.

In the polar coordinate system $(r, \phi)$, when $\k>0$ or $\k=0$,
the range of values of $\Tan_\k(r)$ is the whole real line (completed
with $\infty$), but when $\k<0$, the values of $\Tan_\k(r)$ are confined
to the interval $[0,1/\sqrt{-\k}]$, or to $[-1/\sqrt{-\k},1/\sqrt{-\k}]$
if negative values for $r$ are allowed along the opposite semi-axis according
to the usual practice.

Now, for the conic (\ref{conicEq1}) the minimum value for $r$ happens
when $\phi=0$.  This means that $D$ and $\e$ are independent constants
when $\k\geq0$ but must fulfil the inequality
$0\leq D/(1+\e) < 1/\sqrt{-\k}$ when $\k<0$.

Notice that in (\ref{conicEq1}) the periastron of the orbit is
placed on the semi-axis $\phi=0$ at $r=r_{per}$, and thus $r_{per}$
is related to $D$ and $\e$ by
$$
  \Tan_\k(r_{\rm per}) = \frac{D}{1+\e}
$$
which for any value of $\k$ has always a unique root for $r_{per}$.
Next, let us look for the intersection of the conic with the
line orthogonal to the conic symmetry axis through the focus, which
corresponds to $\phi=\pi/2$. The distance between this intersection
point and the focus is traditionally called {\it semilatus rectum} of
the conic (in the Euclidean case) and will be denoted by $p$.
For $\k=0$ all Euclidean conics (except the limiting double straight
line with $p=\infty$) intersect this semilatus line.
For any $\k\neq0$, $p$ would satisfy
$$
  \Tan_\k(p) = D \,,
$$
but there is a difference between the cases with $\k\geq0$ or $\k\leq0$.
While for $\k>0$, any $D$ will determine a unique $p$, when $\k<0$ 
this equation
will define a real semilatus rectum only when $D< 1/\sqrt{-\k}$, the semilatus
rectum will be formally infinite when $D= 1/\sqrt{-\k}$ and will not exist
at all when $D> 1/\sqrt{-\k}$. This means that in the case of the
hyperbolic plane $H^2_\k$ the family of conics we are considering includes
conics intersecting the `semilatus' line at a proper point, only at
infinity or not intersecting at all. We have arrived to the following
situation: all conics with given fixed focus and symmetry axis intersect this
line in $S^2_\k$, exactly one does not intersect in
$E^2$ and an infinite number of them do not intersect in $H^2_\k$.

Thus, when $\k\geq0$ the equation of the complete family of conics
we are considering is
\begin{equation}
  \Tan_\k(r) = \frac{\Tan_\k(p)}{1+\e \cos\phi}\label{ConicS2}
\end{equation}
but in the negative curvature case $\k<0$, there will be some conics
---precisely those with  $D> 1/\sqrt{-\k}$---
not described under (\ref{ConicS2}). To cater for these cases it will
prove useful to introduce another real distance $\wt{p}$, complementary
to the ideal semilatus rectum, related to $D$ by
$$
  \frac{1}{(-\k) \Tan_\k(\wt{p})}=D
$$
and in terms of this choice, when $\k<0$  the equation of the complete
family of conics we are considering is given by one of the two mutually
exclusive possibilities
\begin{equation}
  \Tan_\k(r) = \frac{\Tan_\k(p)}{1+\e\cos\phi} \,,
  \qquad\qquad
  \Tan_\k(r) = \frac{1}{(-\k)\Tan_\k(\wt{p})(1+\e\cos\phi)} \,,
\label{ConicH2}\end{equation}
where the ranges are $0\leq p <\infty$ and $\infty>\wt{p}\geq0$ respectively.
The conic
\begin{equation}
  \Tan_\k(r) = \frac{1}{\sqrt{-\k}\,(1+\e \cos\phi)} \label{ConicH2Sep}
\end{equation}
is the common limit $p\to\infty$ and $\wt{p}\to\infty$ of (\ref{ConicH2}).
Notice the two expressions (\ref{ConicH2}) can be used as well when $\k>0$ but
then each of the two alternatives covers actually all cases, and are 
then redundant.
Only the first possibility in (\ref{ConicH2}) has a sensible Euclidean limit
because the $\k\to0$ limit of the $\wt{p}$ family gives a straight 
line at the infinity
of the Euclidean plane;
thus the family $\wt{p}$ as a set of conics different from the $p$ family is
specific to the hyperbolic plane.

  Now the only remaining problem is to link the values of the two parameters
$D$ and $\e$ to the type the conic belongs. When is
such a conic an ellipse, a parabola or a hyperbola according to their
definitions in the space of curvature $\k$? When is it a particular conic,
either a circle, a equiparabola or a line?  And finally, ---only for
the hyperbolic plane $H^2_\k$---, when is it a limiting horoellipse or
horohyperbola?

  In the Euclidean plane the answer is well known and easy: the
conic type depends {\it only} on $\e$, and not on
$p$. This is {\it not so} when $\k\neq0$.

\subsection{Conics on the sphere $S^2_\k$}

  A given conic in $S^2_\k$ will have a unique and well defined
parameter $\e$ in (\ref{ConicS2}). But labelling  it as an
ellipse, or parabola or hyperbola, requires to choose a particular
set of two focal elements, and a $\pm$ sign to decide between the
sum or difference in the definitions. Unlike on $H^2_\k$ where
these choices are unique, for the sphere they may be made in
several ways. By means of suitable choices, {\it any  spherical
conic with any $0\leq \e \leq \infty$} may be considered as an
ellipse, as a parabola or as a hyperbola. This is an unavoidable
consequence of the definitions, which are clearly natural ones for
non-zero curvature.

  If we want to classify conics in $S^2$ into three {\it disjoint} species,
this would require adopting an additional convention (somehow bypassing
the definition), which may be chosen in several reasonable ways, but which
nevertheless remains as a convention.

  A possibility would be to class spherical ellipses with $0<e<1$ as `ellipses',
those with $e=1$ as `parabolas', and those with $e>1$ as
`hyperbolas'. This is the convention adopted by Higgs \cite{Hi79}
who formulated it according to the property of not crossing,
touching or crossing the equator, which can be easily shown to be
equivalent  to the requirement $e<1$, $e=1$, or $e>1$,
respectively. Another different convention follows from the fact
that any ellipse/hiperbola with focus $F_1, F_2$ (and focal
separation $2f$) can also be considered as a hyperbola/ellipse
with focus $F_1,\overline{F_2}$ (and focal separation
$2\wt{f}=2(\pi/2\sqrt{\k}-f)$). The sum of the two focal
half-separations $f$ and $\wt{f}$ is $\pi/2\sqrt{\k}$, this is,
the length of a quadrant on the sphere. Thus,  we may consider as
`ellipses' exclusively those spherical ellipses with a value of
the focal separation $2f$ less than a quadrant, that is,
$2f<\pi/2\sqrt{\k}$, and as `hyperbolas' exclusively those
spherical ellipses with a value of $2f$ greater than a quadrant,
that is, $2f>\pi/2\sqrt{\k}$. Within this convention, and if we
want artificially enforce non-redundancy, the name `parabolas'
will only be left for the spherical ellipses with precisely
$2f=\pi/2\sqrt{\k}$, which in our previous nomenclature were
called equiparabola (as $2f=\pi/2\sqrt{\k}$ between two focus
means that one focus will be incident with the the polar of the
other focus, which is the focal line). In this view `ellipses' are
completely contained in the half-sphere with center at the ellipse
center, while `hyperbolas' will go through the boundary of the
half-sphere centered at the hyperbola center.

\subsection{Conics on the hyperbolic plane $H^2_\k$}

  The case of the hyperbolic plane $H^2_\k$ is different and there
is no conventionality in it.  In the following paragraphs we are
implicitly assuming $\k<0$.

  When is the conic (\ref{ConicH2}) an {\it ellipse}? In addition to
the actual vertex at the point of closest approach to the focus,
placed on the $\phi=0$ axis, ellipses will have another vertex,
placed on the $\phi=\pi$ semi-axis with a {\it positive} value for
$\Tan_\k(r_{\rm apo})$ which anyhow should belong to the interval
$[\Tan_\k(r_{\rm peri}), 1/\sqrt{-\k}]$.
The lower bound in this interval corresponds to circular orbits
while the upper places the apoastron at the infinity, and the
ellipse will go to an {\it horoellipse}.
It is easy to conclude that the positivity condition requires
$\e$ to be in the interval $[0,1]$ and then
$\Tan_\k(r_{\rm apo})\in [\Tan_\k(r_{\rm peri}), 1/\sqrt{-\k}]$
only happens within the first alternative in (\ref{ConicH2}),
provided that the eccentricities lie in the interval
$$
  0<\e_{\rm ell}<1-\sqrt{-\k}\,\Tan_\k(p)  \,.
$$
Notice this interval {\it depends on $p$} and for any $p\neq0$ is
strictly smaller than the Euclidean one, which is recovered of course,
when $\k\to 0$. The lower bound for $\e$ corresponds to circular orbits,
while the upper bound is the limiting value corresponding to
{\it horoellipses}
$$
  \e_{\rm cir}=0, \qquad \e_{\rm horoell}=1-\sqrt{-\k}\,\Tan_\k(p) \,.
$$

When will the conic (\ref{ConicH2}) be an {\it hyperbola}? In this case,
in addition to the vertex at the periastron, the conic will have another
vertex point (on the other hyperbola branch, thus not actually on the physical
orbit) on the $\phi=0$ semi-axis, which will appear on the equation as a
{\it negative} value for $\Tan_\k(r_{apo})$ on the $\phi=\pi$ semi-axis.
Hence, the values of $\Tan_\k(r)$ for $\phi=\pi$ must belong to the
interval $[-1/\sqrt{-\k}, -\Tan_\k(r_{\rm per})]$, with the lower value
corresponding to {\it horohyperbolas}, and the upper value to lines.
By using the same strategy, it is easy to conclude that the negativity
condition requires $\e\in[1, \infty]$, and then $\Tan_\k(r_{\rm apo})$
may lie in the required interval in either of the two alternatives
(\ref{ConicH2}), with $\e$ parameters respectively in the intervals
$$
1+\sqrt{-\k}\,\Tan_\k(p)<\e_{\rm hyp}, \qquad
1+\frac{1}{\sqrt{-\k}\,\Tan_\k(\wt{p})}<\e_{\rm hyp}, \qquad
$$
Notice both conditions lead to a common limit when $p, \wt{p} \to\infty$,
namely, $2<\e_{\rm hyperbolas}$ for the conics (\ref{ConicH2Sep}).
Here particular and limiting conics correspond to:
$$
  \e_{\rm horohyp}=1+\sqrt{-\k}\,\Tan_\k(p) \,,\qquad
  \e_{\rm horohyp}=1+\frac{1}{\sqrt{-\k}\,\Tan_\k(\wt{p})} \,, \qquad
  \e_{\rm lines}=\infty \,.
$$

    Once we have characterized the intervals of eccentricity for ellipses
and hyperbolas, the remaining gap correspond to parabolas.
Thus,  within the first family:
$$
  1-\sqrt{-\k}\,\Tan_\k(p)<\e_{\rm par} < 1+\sqrt{-\k}\,\Tan_\k(p)  \,,
$$
and within the second
$$
  1-\frac{1}{\sqrt{-\k}\,\Tan_\k(\wt{p})} < \e_{\rm par}    \,.
$$
The particular parabolas with zero focal separation, this is, the
equiparabolas, corresponds to the values:
$$
  \e_{\rm equipar}=1/\Cos_k(p)=\sqrt{1+\k\Tan^2_\k(p)}, \qquad
  \e_{\rm equipar}=\Cos_k(\wt{p})=1/\sqrt{1+\k\Tan^2_\k(\wt{p})}  \,.
$$
and are different from the parabolas with $e=1$.

   This way we get the following result:  conics (\ref{ConicH2}) in 
the Hyperbolic
plane $H^2$, with a fixed value of $p$ may be either ellipses, 
parabolas or hyperbolas,
according to the value of their  eccentricity as follows
\begin{eqnarray}
  &&{\rm  circle} {\hskip 70pt}
    \e_{\rm cir}\ =\ 0        \cr
  &&{\rm  ellipses}{\hskip 60pt}
    0 <\ \e_{\rm ell}\ <\ 1-\sqrt{-\k}\,\Tan_\k(p)  \cr
  &&{\rm  horoellipse}{\hskip 40pt}
    \e_{\rm horoell}\ =\   1-\sqrt{-\k}\,\Tan_\k(p)   \cr
  &&{\rm  parabolas}{\hskip 44pt}
    1-\sqrt{-\k}\,\Tan_\k(p)\ <\ \e_{\rm par}\ <\ 1+\sqrt{-\k}\,\Tan_\k(p)  \cr
  &&{\rm  horohyperbolas}{\hskip 18pt}
    \e_{\rm horohyp}\  =\  1+\sqrt{-\k}\,\Tan_\k(p)   \cr
  &&{\rm  hyperbolas}{\hskip 40pt}
    1+\sqrt{-\k}\,\Tan_\k(p)\ <\ \e_{\rm hyp}\ <\ \infty
{\nonumber}
\end{eqnarray}
and conics (\ref{ConicH2}) with a fixed $\wt{p}$ may be only
parabolas or hyperbolas:
$$
  1-\frac{1}{\sqrt{-\k}\,\Tan_\k(\wt{p})}< \e_{\rm par} <
  1+\frac{1}{\sqrt{-\k}\,\Tan_\k(\wt{p})}< \e_{\rm hyp}<\infty  \,.
$$
The most remarkable property is the existence, for a fixed value of the
semilatus rectum $p$, of a full interval of values for eccentricity
corresponding to parabolas, including always two special values,
$\e=1/\Cos_k(p)$ and $\e=1$; in the Euclidean limit $\k\to 0$ this
interval collapses to the single value $\e_{par}=1$, and there is a
single Euclidean parabola for a given $p$.

    Now let us consider the complete family of conics corresponding to
trajectories with a given {\it periastron} distance, say $r_{\rm per}$;
these would correspond to the trajectories of a particle launched orthogonally
to the radial direction from a point at distance $r_{per}$ from the
potential origin and with a given velocity.  The corresponding pattern
for the Kepler problem in the Euclidean case is well known: a circle, a
family of ellipses of increasing eccentricity $0<\e<1$, a separating
parabola, a family of hyperbolas of increasing eccentricity $1<\e<\infty$
and finally a limiting straight line corresponding to infinite velocity.
How about the analogue of this pattern in the non-zero curvature case?

    First we require all conics will have a periastron at $r_{\rm per}$.
Within the two families the relation between $p$ or $\wt{p}$ with
$r_{\rm per}$ is given by
$$
  \Tan_\k(r_{\rm per}) = \frac{\Tan_\k(p)}{1+\e}
  \qquad\qquad
  \Tan_\k(r_{\rm per}) = \frac{1}{(-\k)\Tan_\k(\wt{p})(1+\e)}
$$
and, using these expression in the classification stated above, we get
the landmark values of $p$ or $\wt{p}$ corresponding to the different
types of conic:
$$
\Tan_\k(r_{\rm per}) <
\Tan_\k(p_{\rm ell}) <
\frac{2\Tan_\k(r_{\rm per})}{1+\sqrt{-\k}\Tan_\k(r_{\rm per})}<
\Tan_\k(p_{\rm par}) <
\frac{2\Tan_\k(r_{\rm per})}{1-\sqrt{-\k}\Tan_\k(r_{\rm per})}<
\Tan_\k(p_{\rm hyp})   \,,
$$
which in the Euclidean limit, with $\k=0$ and $\Tan_0(x)=x$, reduces to
the well known:
$$
  r_{\rm per}< p_{\rm ell}<  2r_{\rm per}=p_{\rm par}=2r_{\rm per}<
  p_{\rm hyp},
\quad\hbox{when}\quad  \k=0   \,.
$$
For the hyperbolic $\wt{p}$ family, which has no Euclidean limit, we
have:
$$
  \Tan_\k(\wt{p}_{\rm hyp}) <
  \frac{1-\sqrt{-\k}\,\Tan_\k(r_{\rm per})}{(- 2 \k) \Tan_\k(r_{\rm per})} <
  \Tan_\k(\wt{p}_{\rm par})    \,.
$$
Figures 6A and 6B represent a set of Kepler orbits in the
hyperbolic plane $H^2$.

\section{Final comments and outlook}

   We have solved the Kepler problem on the three spaces of constant curvature.
As we have stated in the introduction, one of the fundamental characteristic
of this approach is the use of the curvature $\k$ as a parameter.
In this way,  all the $\k$-dependent properties that we have obtained
reduce to the appropriate property for the system on the sphere $S^2$,
or the on hyperbolic plane $H^2$, when particularized for $\k>0$, or $\k<0$,
respectively; in addition, the Euclidean case arises as the very particular
(but important) case $\k=0$.
So, we can summarize this situation pointing out two important facts:
\begin{itemize}
\item  The Kepler problem is not a specific or special characteristic of
the Euclidean space but it is well defined in all the three spaces of
constant curvature.
\item  There are not three different Kepler problems but only one
that is defined, at the same time, in three different manifolds.
\end{itemize}
Of course, since the three manifolds are geometrically different, many
dynamical properties show differences according to the characteristics
of the manifolds;
nevertheless, the important point is that there is only one theory
that is simultaneously valid for the three manifolds and for any value
of the curvature.
We illustrate this situation recalling the following two important
results:
\begin{itemize}
\item  The equation of Binet remains true in the three cases.
\item  The orbits are conics in the three cases.
\end{itemize}

  It is well known that the classical Euclidean Kepler problem is one of
the systems endowed with more interesting properties.
Therefore, it would be interesting to study, all of them, in the general
non-Euclidean $\k\ne 0$ case;  as an example, we have made use, in Sec. 4,
of two constants of motion, $I_3(\k)$ and $I_4(\k)$, that represent
the $\k$-dependent version of the Runge-Lenz vector.
The Runge-Lenz vector plays a very important role in the Euclidean case,
so it is natural to suppose that the same situation will be true for the
case of the general `curved' system.
In fact, we think that, as the Euclidean system is just a very particular
case of a much more general system, all the Euclidean characteristics,
that are obtained and discussed in the books of theoretical mechanics,
must admit a $\k$-dependent deformed version appropriate for the general
``curved" system.
We think that these questions are open problems that must be investigated.
We also note that in all this paper we have only dealt with the
two-dimensional case since we assumed that it arises from a
reduction of the three-dimensional Kepler problem on
$S^3$ and $H^3$;
this dimensional reduction is worthy of a further study.

  Finally, the analysis of the orbits has led us, in a natural way,
to the theory of conics on spaces of constant curvature.
Notice that, although Secs. 3 and 4 were mainly concerned with
dynamical questions, Sec. 5 was written emphasizing its geometrical
character; in fact, it can be considered by itself and independently
of the other previous sections.
It is clear that the theory of conics on spaces of constant curvature
is a geometrical matter of great importance deserving a deeper study
that we hope to present elsewhere.
On the one hand, there exist some general points and basic properties
that, although perfectly known and clearly stated in the Euclidean plane,
remain to be studied in the case of spaces of constant curvature;
on the other hand, we have already obtained some particular points as,
for example, the existence several different parabolas in the Hyperbolic
plane, or the possibility of some different alternative ways of defining
the eccentricity for a conic in a $\k\ne 0$ space, which are really
noteworthy.
These geometrical questions are also open problems to be investigated.

\section*{Appendix: Polar geodesic coordinates}

    A two--dimensional manifold $M$ can be described by using different
coordinate systems.
If we consider it as an imbedded submanifold of $\IR^3$, then the points
of $M$ can be characterized by the three external coordinates,
as $(x,y,z)$ or $(r,\phi,\te)$, plus an additional constraint relation.
Nevertheless, in differential geometric terms, a more appropriate approach
is to develop the study by using two--dimensional systems of coordinates
adapted to $M$.

    On any general two--dimensional Riemannian space,
not necessarily of constant curvature, there are two distinguished
types of local coordinate systems: ``geodesic parallel" and
``geo\-desic polar" coordinates. They reduce to the familiar
cartesian and polar coordinates on the Euclidean plane (see Refs.
\cite{RaS99} and \cite{Kl78}) and both are based on a origin point
$O$ and  a oriented geodesic $g_1$ through $O$.

   For any point $P$ in some suitable neighbourhood of $O$, there is a
unique geodesic $g$ joining $P$ with $O$. The (geodesic) polar
coordinates $(r,\phi)$ of $P$, relative to the origin $O$ and the
positive geodesic ray of $g_1$, are the (positive) distance $r$
between $P$ and $O$ measured along $g$, and the angle $\phi$
between $g$ and the positive ray $g_1$, measured around $O$ (Fig.
7). These coordinates are defined in a neighbourhood of $O$ not
extending beyond the cut locus of $O$; polar coordinates are
singular at $O$, and $\phi$ is discontinuous on the positive ray
of $g_1$.

   In the case of $M$ being a space of constant curvature $\k$,
the expression for the differential element of distance $ds^2$
is given by
$$
   ds_\k^2 = d\,r^2 + \Sin_\k^2(r)\,d{\phi}^2 \,,
$$
so that we get $ds^2 = d\,r^2 + r^2\,d{\phi}^2$ for the
particular $\k=0$ Euclidean case.

{\small
\section*{\bf Acknowledgments.}
Support of projects  BFM-2003-02532, FPA-2003-02948,
BFM-2002-03773, and CO2-399 is acknowledged.

 }

\vfill\eject
\section*{Figure Captions}

\begin{itemize}
\item{} {\sc Figure 1}.{\enskip}
Plot of the Kepler Potential as a function of $r$, for the unit
sphere $\k=1$ (upper curve), Euclidean plane $\k=0$ (dash line),
and `unit` Lobachewski plane $\k=-1$ (lower curve).
The three functions are singular at $r=0$ but the Euclidean function
$U_0=V$ appears in this formalism as making a separation between
two different behaviours. In fact $U_0=V$ is the only Potential
that vanish at long distances.

\item{} {\sc Figure 2}.{\enskip} Plot of $W_c$ as a function of
$r$, for $\k=1$ with $k$ and $J$ given by ($k=1, J=1$).

\item{} {\sc Figure 3}.{\enskip} Plot of $W_c$ as a function of
$r$, for ($k=1, J=1$) and three different values of the curvature
$c=\sqrt{\k}$ : $c=1.2$ (upper curve), $c=0.8$ (middle curve), and
$c=0.4$ (lower curve).

\item{} {\sc Figure 4}.{\enskip} Plot of $W_{-c}$ as a function
of $r$, for $c=1$ ($\k=-1$). The upper curve corresponds to ($k=1,
J=2$) and lower curve corresponds to ($k=4, J=1$).

\item{} {\sc Figure 5}.{\enskip} Plot of $W_{-c}$ as a function
of $r$, for ($k=4, J=1$) and three different values of the
curvature $c=\sqrt{-\k}$ : $c=1.5$ (lower curve), $c=1$ (middle
curve), and $c=0.5$ (upper  curve).

\item{} {\sc Figure 6A and 6B}.{\enskip} A set of Kepler orbits
with a fixed periastron distance, depicted in the conformal
Poincare disk model of hyperbolic plane, for `small' $r_{\rm per}$
(Figure 6A)  and `large' $r_{\rm per}$ (Figure 6B). The potential
center is at the origin, which is a focus of the conics. Thick
lines represent particular and limiting conics:  circle,
horoellipse, horohyperbola and straight line. A suitable selection
(with the other  focal elements chosen as to make the diagram
clear) of ellipses, parabolas and hyperbolas are represented as
thin lines in each of the three ranges determined by the  previous
conics. For ellipses and hiperbolas the other focus (not marked)
is on the horizontal line; for parabolas the focal line is
perpendicular to this horizontal line. The  semilatus rectum of
the conic lies on the vertical straight line through the origin;
notice in the figure (A) only some hyperbolas do not intersect
this line, while in (B) conics  which do not intersect the
semilatus line include all hyperbolas as well as many  parabolas;
these are the two generic behaviours, as explained in the text.

\item{} {\sc Figure 7}.{\enskip} Polar geodesic coordinates in a
two--dimensional Riemannian space $(M,g)$ (Figure 7A) and in the
sphere $S^2$ (Figure 7B).
\end{itemize}

\vfill\eject\null

{\vskip 30pt}
$$
\epsfbox{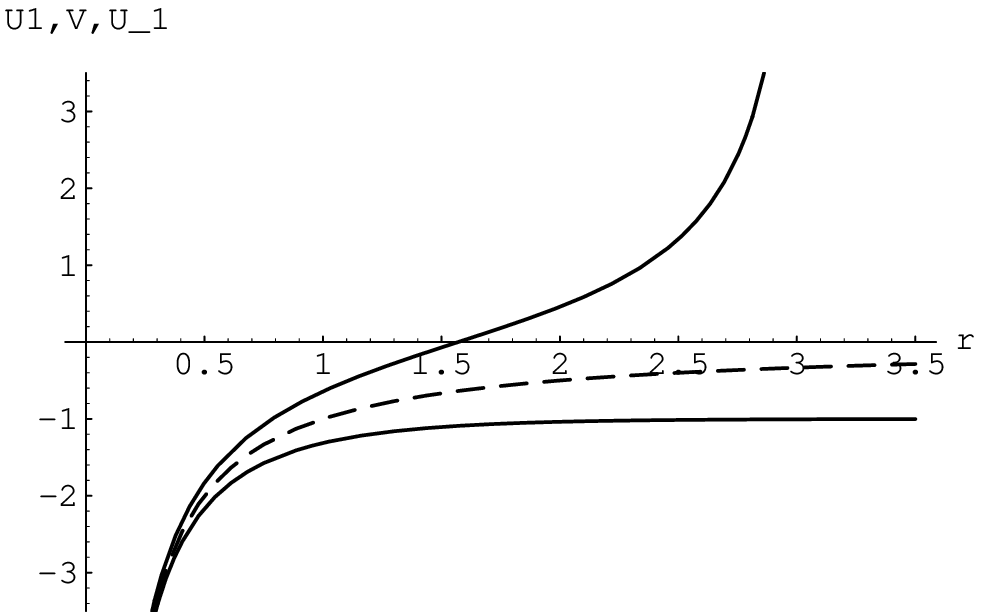}
$$
{\smallskip}

{\sc Figure 1}.{\enskip} Plot of the Kepler Potential
as a function of $r$, for the unit sphere $\k=1$ (upper curve),
Euclidean plane $\k=0$ (dash line), and `unit` Lobachewski plane
$\k=-1$ (lower curve). The three functions are singular at $r=0$
but the Euclidean function $U_0=V$ appears in this formalism as
making a separation between two different behaviours. In fact
$U_0=V$ is the only Potential that vanish at long distances.

\vfill\eject\null
$$
\epsfbox{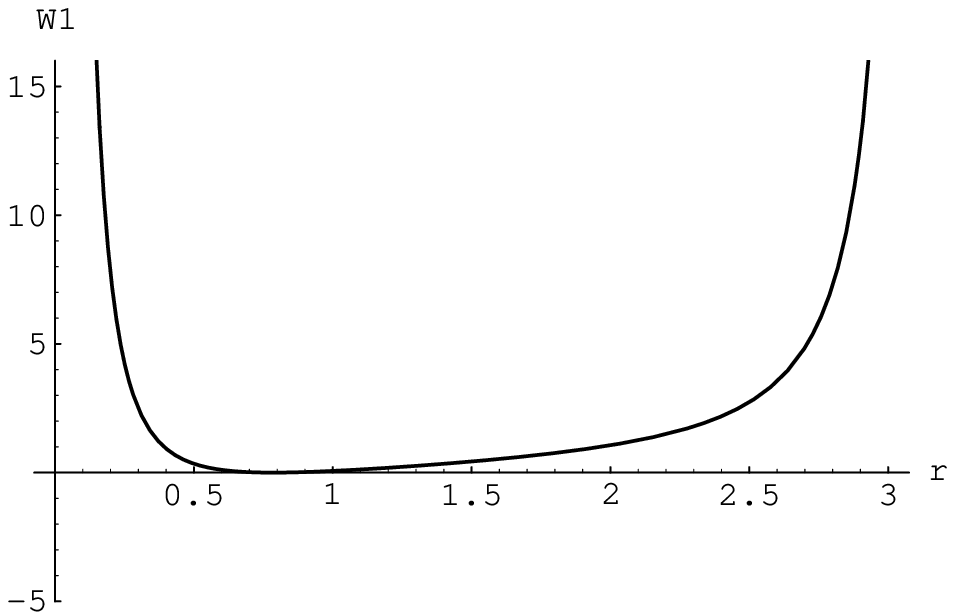}
$$
{\smallskip}

{\sc Figure 2}.{\enskip} Plot of $W_c$ as a function of
$r$, for $\k=1$ with $k$ and $J$ given by ($k=1, J=1$).

{\vskip 30pt}
$$
\epsfbox{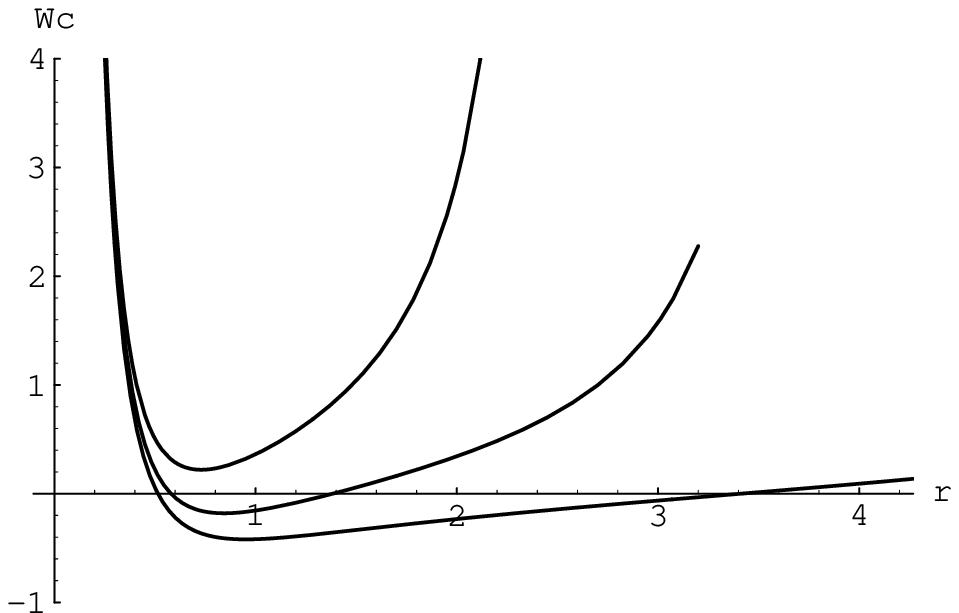}
$$
{\smallskip}

{\sc Figure 3}.{\enskip} Plot of $W_c$ as a function of
$r$, for ($k=1, J=1$) and three different values of the curvature
$c=\sqrt{\k}$ : $c=1.2$ (upper curve), $c=0.8$ (middle curve), and
$c=0.4$ (lower curve).

\vfill\eject\null
$$
\epsfbox{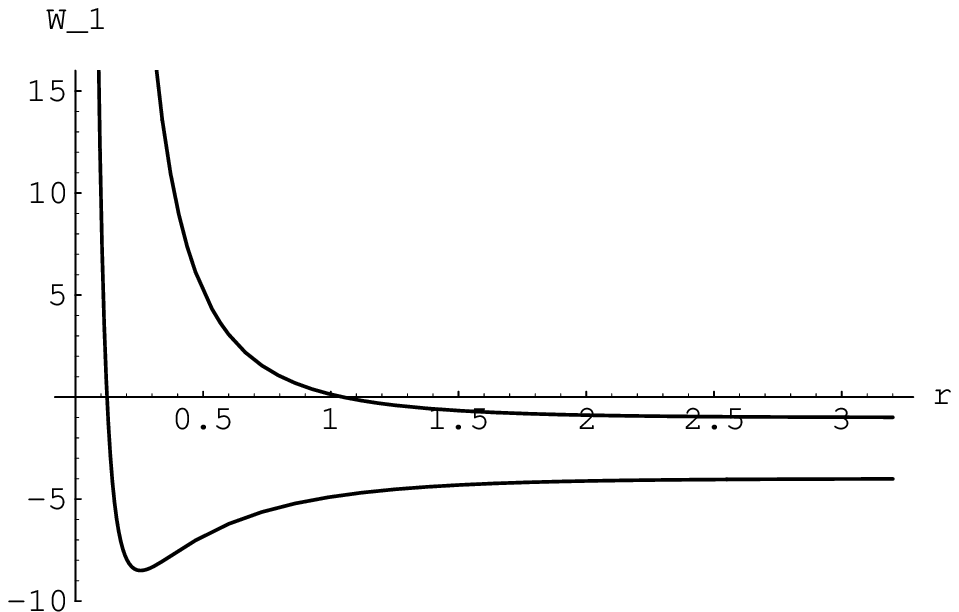}
$$
{\smallskip}

{\sc Figure 4}.{\enskip} Plot of $W_{-c}$ as a function
of $r$, for $c=1$ ($\k=-1$). The upper curve corresponds to ($k=1,
J=2$) and lower curve corresponds to ($k=4, J=1$).

{\vskip 30pt}
$$
  \epsfbox{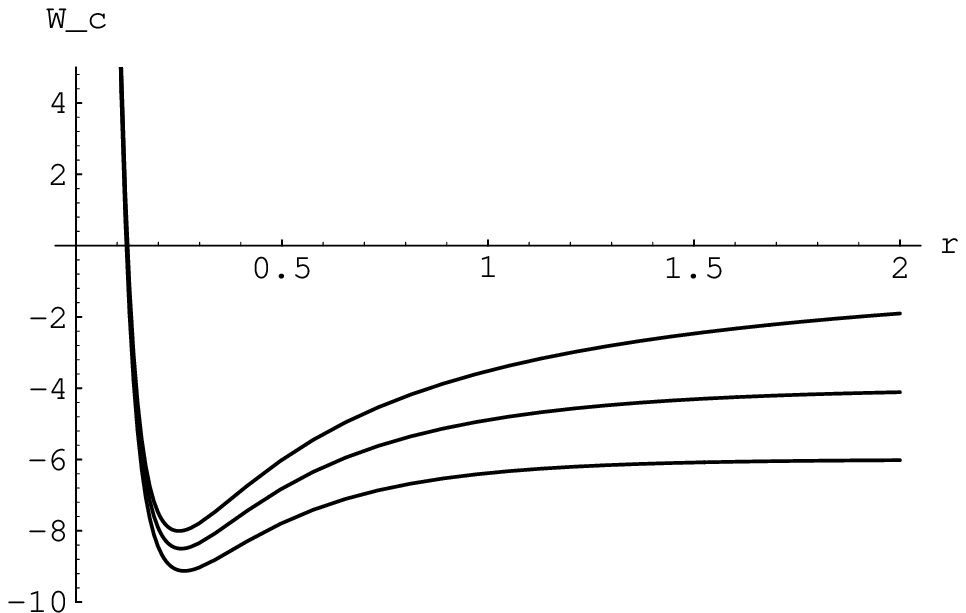}
$$
{\smallskip}

{\sc Figure 5}.{\enskip} Plot of $W_{-c}$ as a function
of $r$, for ($k=4, J=1$) and three different values of the
curvature $c=\sqrt{-\k}$ : $c=1.5$ (lower curve), $c=1$ (middle
curve), and $c=0.5$ (upper  curve).

\vfill\eject
\epsfbox{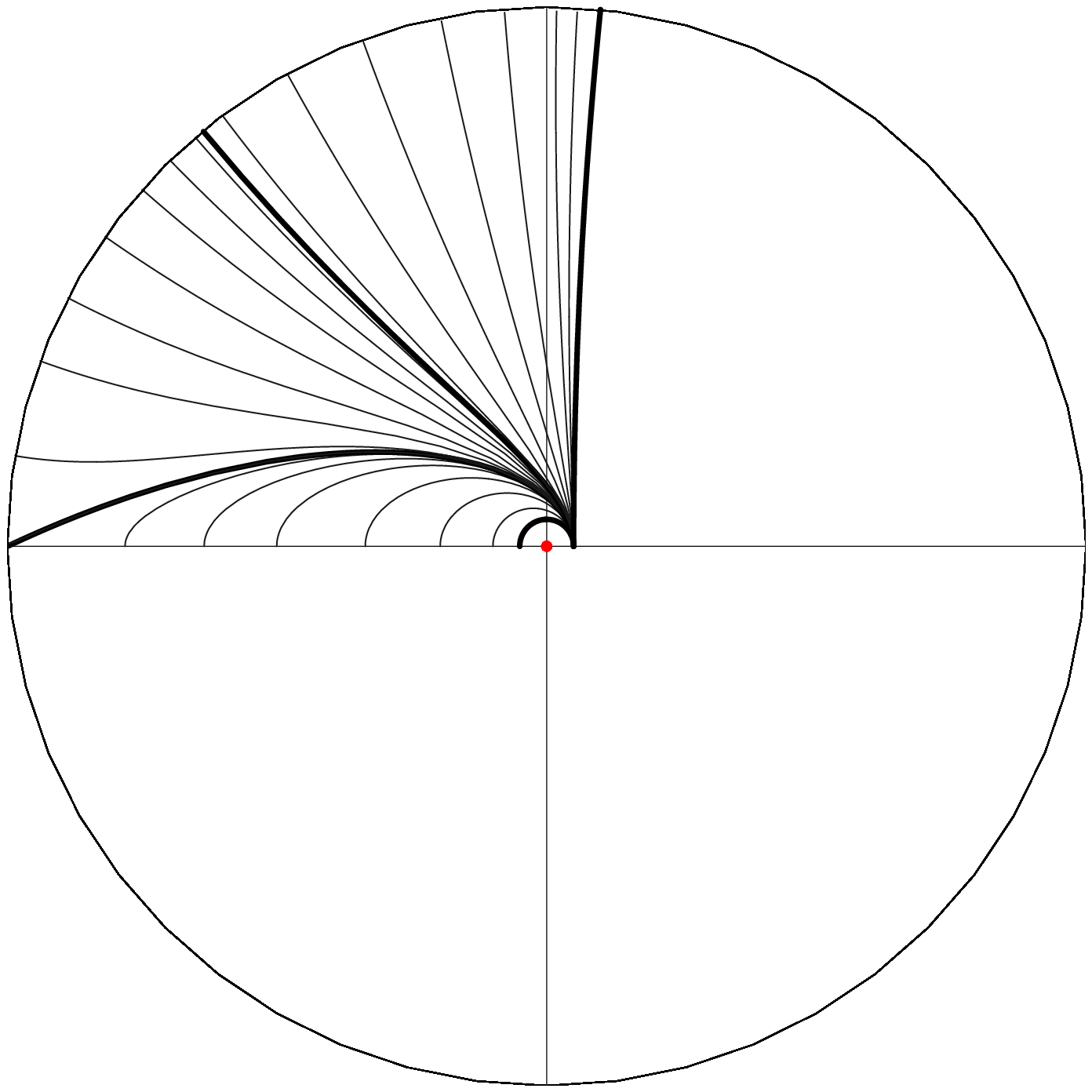}

{\sc Figure 6A and 6B}.{\enskip} A set of Kepler orbits with a
fixed periastron distance, depicted in the conformal Poincare disk
model of hyperbolic plane, for `small' $r_{\rm per}$ (Figure 6A)
and `large' $r_{\rm per}$ (Figure 6B). The potential center is at
the origin, which is a focus of the conics. Thick lines represent
particular and limiting conics:  circle, horoellipse,
horohyperbola and straight line. A suitable selection (with the
other  focal elements chosen as to make the diagram clear) of
ellipses, parabolas and hyperbolas are represented as thin lines
in each of the three ranges determined by the  previous conics.
For ellipses and hiperbolas the other focus (not marked) is on the
horizontal line; for parabolas the focal line is perpendicular to
this horizontal line. The  semilatus rectum of the conic lies on
the vertical straight line through the origin;  notice in the
figure (A) only some hyperbolas do not intersect this line, while
in (B) conics  which do not intersect the semilatus line include
all hyperbolas as well as many  parabolas; these are the two
generic behaviours, as explained in the text.

\vfill\eject
\epsfbox{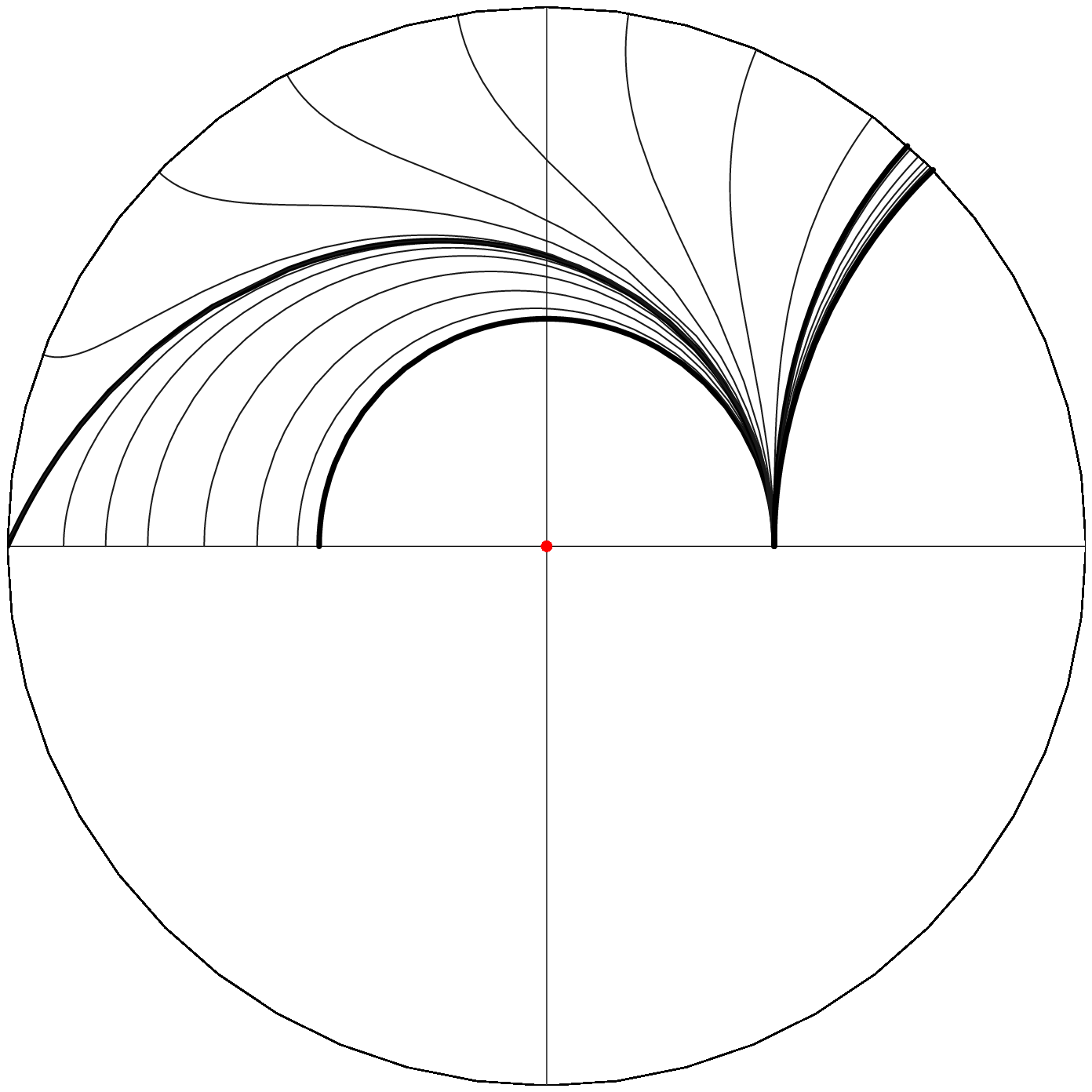}

{\sc Figure 6A and 6B}.{\enskip} A set of Kepler orbits with a
fixed periastron distance, depicted in the conformal Poincare disk
model of hyperbolic plane, for `small' $r_{\rm per}$ (Figure 6A)
and `large' $r_{\rm per}$ (Figure 6B). The potential center is at
the origin, which is a focus of the conics. Thick lines represent
particular and limiting conics:  circle, horoellipse,
horohyperbola and straight line. A suitable selection (with the
other  focal elements chosen as to make the diagram clear) of
ellipses, parabolas and hyperbolas are represented as thin lines
in each of the three ranges determined by the  previous conics.
For ellipses and hiperbolas the other focus (not marked) is on the
horizontal line; for parabolas the focal line is perpendicular to
this horizontal line. The  semilatus rectum of the conic lies on
the vertical straight line through the origin;  notice in the
figure (A) only some hyperbolas do not intersect this line, while
in (B) conics  which do not intersect the semilatus line include
all hyperbolas as well as many  parabolas; these are the two
generic behaviours, as explained in the text.

\vfill\eject

$$
  \epsfbox{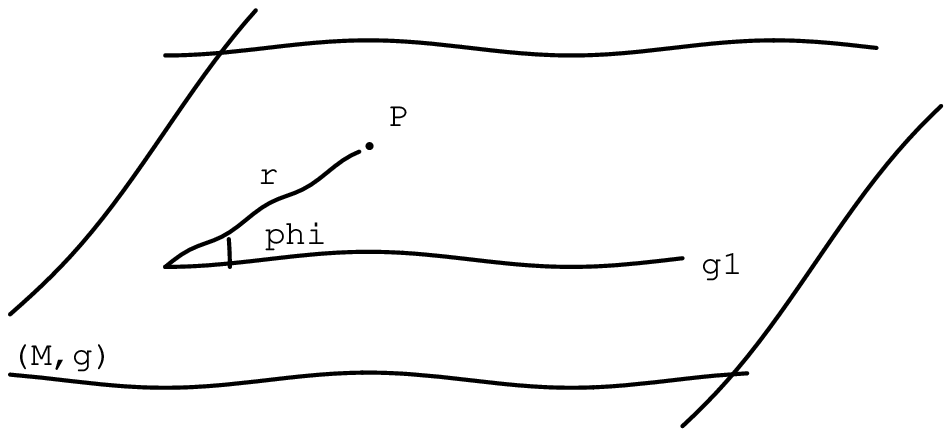}
$$
{\smallskip}

{\sc Figure 7A}.{\enskip}  Polar geodesic coordinates in a
two--dimensional Riemannian space $(M,g)$.


$$
  \epsfbox{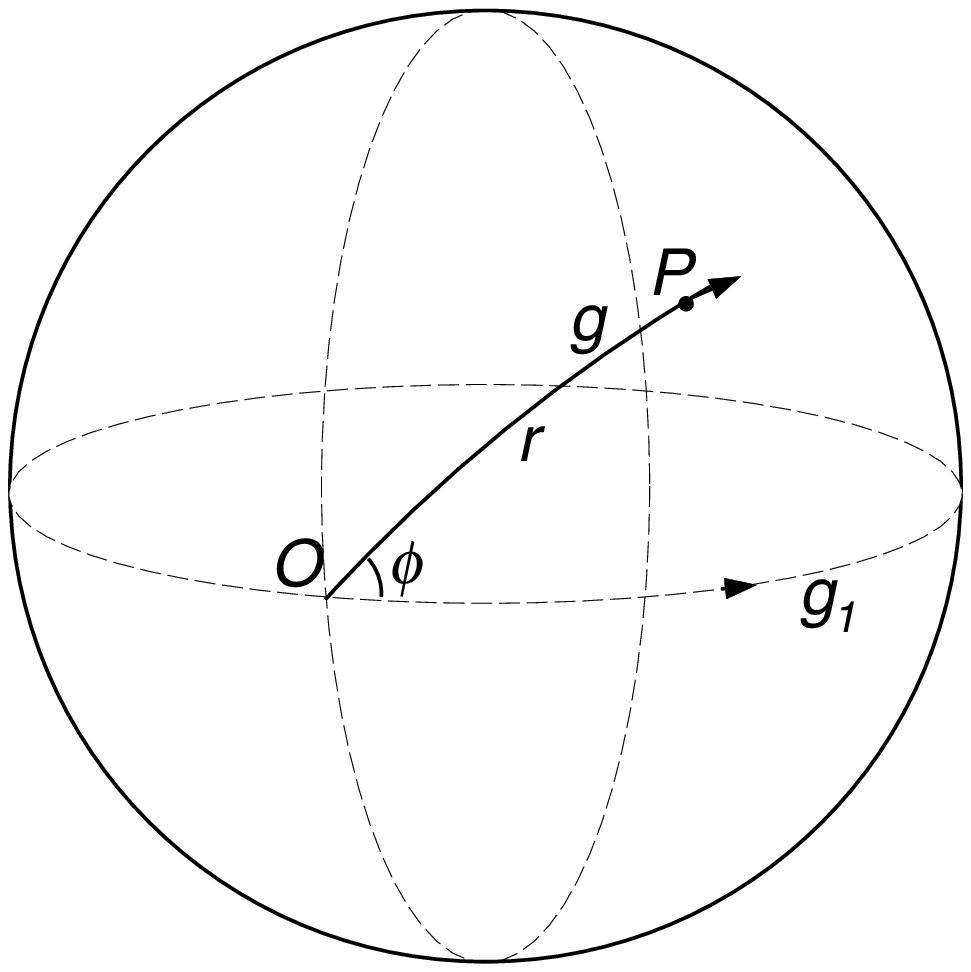}
$$


\end{document}